\documentclass[pra,floatfix,nofootinbib,reprint,superscriptaddress]{revtex4-1}

\usepackage{setstack,amsopn} 
\expandafter\let\csname equation*\endcsname\relax
\expandafter\let\csname endequation*\endcsname\relax
\usepackage{amsmath, amssymb, amsthm}
\usepackage{dcolumn}
\usepackage{dutchcal}
\usepackage{mathbbol}
\usepackage{tikz} 
\usepackage[english]{babel}
\usepackage{times}
\usepackage{tabularx}
\usepackage{graphicx}
\usepackage{amsfonts}
\usepackage{siunitx}
\usepackage{mathtools}
\usepackage{stmaryrd}
\usepackage[e]{esvect}
\usepackage{tikz}
\usepackage{tablefootnote}
\usetikzlibrary{matrix}
\usepackage[pdftex, colorlinks, allcolors=blue]{hyperref}
\usepackage{cleveref}
\DeclareSymbolFont{sfletters}{OML}{cmbrm}{m}{it}
\DeclareMathOperator{\cc}{c.c.}

\DeclareMathOperator{\Real}{Re}

\newcommand{\dket}[1]{|#1\rangle}
\newcommand{\dbra}[1]{\langle#1|}

\newcommand\todo[1]{\textcolor{blue}{#1}}

\newcommand{\norm}[1]{\left|\left|#1\right|\right|}
\newcommand{\parenthesis}[1]{\left(#1\right)}

\makeatother
\usepackage{algorithm}
\usepackage[noend]{algpseudocode} 
\begin{document}

\newcommand{\overbar}[1]{\mkern 1.5mu\overline{\mkern-1.5mu#1\mkern-1.5mu}\mkern 1.5mu}

\title{Optimal control of large quantum systems: assessing memory and runtime performance of GRAPE}

\author{Yunwei Lu}\email{yunweilu2020@u.northwestern.edu}
\author{Sandeep Joshi}
\author{Vinh San Dinh}
\author{Jens Koch}
\address{Department of Physics and Astronomy, Northwestern University, Evanston, Illinois 60208, USA}

\date{\today}

\begin{abstract}
Gradient Ascent Pulse Engineering (GRAPE) is a popular technique in quantum optimal control, and can be combined with automatic differentiation (AD) to facilitate on-the-fly  evaluation of cost-function gradients. We illustrate that the convenience of AD comes at a significant memory cost due to the cumulative storage of a large number of states and propagators. For quantum systems of increasing Hilbert space size, this imposes a significant bottleneck. We revisit the strategy of hard-coding gradients in a scheme that fully avoids propagator storage and significantly reduces memory requirements. Separately, we present  improvements to numerical state propagation to enhance runtime performance. We benchmark runtime and memory usage and compare this approach to AD-based implementations, with a focus on pushing towards larger Hilbert space sizes. The results confirm that the AD-free  approach facilitates the application of optimal control for large quantum systems which would otherwise be difficult to tackle.
\end{abstract}

\maketitle

\section{Introduction}
The rapidly evolving field of quantum information processing has generated much interest in quantum optimal control for tasks such as state preparation  \cite{rojan2014arbitrarystate}, error correction \cite{waldherr2014quantumerror,gertler2021protecting} and realization of logical gates \cite{scabdelhafez2020universal,scallen2017optimal,schuang2014optimal,scwerninghaus2021leakage}. This methodology has been implemented in various quantum systems \cite{glaser2015training}, such as nuclear magnetic resonance \cite{nmrbretschneider2012conversion,nmrkehlet2004improving,nmrkhaneja2005optimal,nmrsimon2009optimal,nmrspinachhogben2011spinach,nmrxu2008designing}, trapped ions \cite{timuller2015phonon,tinebendahl2009optimal}, nitrogen-vacancy centers in diamonds \cite{nvangerer2013robust,nvchou2015optimal,nvdolde2014high,nvpoggiali2018optimal,nvpoulsen2021optimal,nvrembold2020introduction,nvyan2021population}, Bose-Einstein condensation  \cite{becamri2019optimal,becamri2020optimal,becmennemann2015optimal,becsorensen2018qengine} and neutral atoms \cite{atomguo2019high,atomrosi2013fast}. 

One of the prominent techniques for implementing quantum optimal control is Gradient Ascent Pulse Engineering (GRAPE). The central goal of optimal control is to adjust control parameters in such a way that the infidelity of the desired state transfer or gate operation is minimized. In GRAPE, the minimization is based on gradient descent and thus generally requires the evaluation of gradients. Automatic differentiation (AD) \cite{baydin2018automatic} is a convenient tool that has also made an impact on quantum optimal control with GRAPE \cite{pacleung2017speedup,pacqoc,abdelhafez2019gradient}.
Internally, AD builds a computational graph and applies the chain rule to automatically compute the gradient of a given function. 

However, the convenience of AD comes at the cost of memory required for storing the full computational graph. The use of semi-automatic differentiation (semi-AD) \cite{semiadgoerz2022quantum} partially reduces the memory overhead compared to full automatic differentiation (full-AD). This reduction can still be insufficient to tackle optimal control for quantum systems of rapidly growing size \cite{arute2019quantum,gong2021quantum}. This motivates us to revisit the original GRAPE scheme \cite{nmrkhaneja2005optimal} which is based on hard-coded gradients (HG), and has the smallest possible memory cost. We perform a scaling analysis of memory usage and runtime, and thereby develop a decision tree guiding the optimal choice of strategy among HG, semi-AD and full-AD. We exemplify the scaling behavior with benchmarks for concrete optimization tasks.

The outline of our paper is as follows. Section \ref{grapetheory} briefly reviews GRAPE and introduces necessary notation. In Sec.\ \ref{gradients main text}, we illustrate derivation and memory-efficient implementation of the gradients for the most commonly used cost function contributions. Our numerical implementation of HG further improves the efficiency of state propagation. In Sec.\ \ref{Scaling analysis and benchmarking}, we analyze and compare the scaling of runtime and memory usage scaling for HG, AD and semi-AD, and present results from concrete benchmark studies. In Sec.\ \ref{Decision tree}, we formulate a decision tree that facilitates the choice of optimal numerical strategy for GRAPE-based quantum optimal control. We summarize our findings and present our conclusions in Sec.\ \ref{Conclusions}.

\section{Brief sketch of GRAPE \label{grapetheory}}

We briefly sketch the basics of GRAPE \cite{nmrkhaneja2005optimal}, mainly to establish the notion used in subsequent sections. 

Consider a system described by the Hamiltonian
\begin{align}
     H(t) = H_s + a(t)\, h_c.
\end{align}
Here, $H_s$ denotes the static system Hamiltonian and $h_c$ is an operator coupling the classical control field $a(t)$ to the system\footnote{For simplicity, we consider one control channel with real-valued $a(t)$. Generalizations to multiple control channels and complex $a$ are straightforward.}. Quantum optimal control aims to adjust $a(t)$ such that a desired unitary gate or state transfer is performed within a given time interval $0\le t \le T$. To facilitate optimization, the total control time $T$ is divided into $N$ intervals of duration $\Delta t=T/N$, and the control field is specified by the amplitudes at the discrete times $t_n=n\Delta t$ ($n=1,2,\dots,N$). The discretization interval $\Delta t$ is chosen such that control amplitudes are approximately constant\footnote{This treatment is actually quite close to modeling the actual output of an arbitrary waveform generator, where $\Delta t$ can be chosen to align with the available resolution.} within each $\Delta t$. We denote the time-discretized values by
\begin{align}
\label{eq2}
    a_n \coloneqq a(t_n),
\end{align}
and group these to form the vector $\textbf{a}\in \mathbb{R}^{N}$.  
Under the influence of this piecewise-constant control field, the system's quantum state $\dket{\psi_n}\coloneqq \dket{\psi(t_n)}$ evolves by a single time step according to 
\begin{align}
\label{statepropagation}
  |\psi_n\rangle=U_n|\psi_{n-1}\rangle.
\end{align}
Here, $U_n$ is the short-time propagator
\begin{align}
\label{propagator}
  U_n\coloneqq U(t_n, t_{n-1}) =e^{-iH_n\Delta t}\qquad(\hbar=1),
\end{align}
where $H_n=-i\, (H_s+a_n\, h_c )\Delta t$. Optimization of the control field proceeds via minimization of a cost function $C$. Typically, $C$ includes the state-transfer or gate infidelity, along with additional cost contributions employed to moderate pulse characteristics such as maximal power or bandwidth. To this end a composite cost function is constructed, $C(\textbf{a})=\sum_{\nu}\alpha_\nu C_\nu \parenthesis{\textbf{a}}$, where $\alpha_\nu$ are empirically chosen weight factors and $C_\nu$ are individual cost contributions. GRAPE utilizes the method of steepest descent to minimize $C(\textbf{a})$ by updating the control amplitudes according to the cost function gradient: 
\begin{align}
  \textbf{a}^{(j+1)}=\textbf{a}^{(j)}-\eta_j\,\nabla_{\textbf{a}}C(\textbf{a}^{(j)}).
\end{align}
Here, $j$ enumerates steepest-descent iterations and $\eta_j>0$ is the learning rate governing the step size. 

\section{Calculation of gradients}
\label{gradients main text}
A critical ingredient for GRAPE is the computation of cost function gradients. One popular option is to leverage automatic differentiation for this purpose, such as implemented in \texttt{Tensorflow} \cite{tensorflow2015-whitepaper} and \texttt{Autograd} \cite{maclaurin2015autograd}. A clear advantage of this strategy is the ease by which complicated cost function gradients are evaluated automatically at runtime. This convenience, however, is bought at the cost of significant memory consumption which can be prohibitive for large quantum systems. 
This waste of memory resources is prohibitive when optimal control is applied to large quantum systems.
Therefore, we are motivated to revisit hard-coded gradients in a computationally efficient scheme for key cost contributions.

\subsection{Analytical gradients \label{c1c2}}
We provide a discussion of the analytical gradients of two key cost contributions in this subsection. Expressions for gradients of other cost contributions are shown in App.\  \ref{gradient}.

In state-transfer problems, a central goal of quantum optimal control is to minimize the state-transfer infidelity given by 
\begin{equation}
\label{C1st}
    C_1^
{st}=1-\left|\left\langle\phi_T | \psi_{N}\right\rangle\right|^{2},
\end{equation} 
where $\dket{\phi_T}$ is the target state and $\dket{\psi_N}$ is a state realized at final time $t_N$. The gradient of $C_1^{\text{st}}$ takes the form
\begin{align}
\nonumber \frac{\partial C_1^{st} }{\partial a_{ n}}
  &=- \langle\phi_T|U_N\cdots \frac {\partial U_{n}}{\partial a_{n}}\cdots U_1|\psi_0\rangle\langle\psi_N|\phi_T\rangle -\cc\\
\label{stgradient}  &=-\langle\phi_{n}|\frac {\partial U_n }{\partial a_{n}}|\psi_{n-1}\rangle\langle\psi_N|\phi_T\rangle - \cc ,
\end{align}
where the state $\dket{\phi_{n}}$ obeys the recursive relation
\begin{equation}
    \dket{\phi_{n}}=U_{n+1}^\dagger\dket{\phi_{n+1}}.
\end{equation}
This can be interpreted as backward propagation.

In addition to state transfer, unitary gates are another operation of interest. Gate optimization can be achieved by minimizing the gate infidelity $C_1^{g}=1-\big|\operatorname{tr}\big(U_T^{\dagger} U_R\big)/d\big|^2$. Here, $U_T$ is the target unitary gate and $U_R=U_N\cdots U_1$ is the actually realized gate. For a given  orthonormal basis $\{\dket{\psi^h_0}\}_h$ ($h$ enumerates the basis states), the gradient of $C_1^{ g}$ can be written as 

\begin{equation} 
\begin{aligned}
\label{gategradient}
  \frac{\partial C_1^{ g} }{\partial a_{ n}}
  &=-\frac{1}{d^2}\sum_{h}\langle\phi_n^h|\frac {\partial  U_n}{\partial a_{n}}|\psi^h_{n-1}\rangle \sum_{h'}\dbra{\psi_N^{h'}}\phi_N^{h'}\rangle
  -\cc,\\
\end{aligned}
\end{equation}
where $\dket{\psi_{n}^h}=U_n\cdots U_1\dket{\psi_0^h}$. The state $\dket{\phi_{n}^h}$ is obtained as follows: apply the target unitary $U_T$ to basis state $\{\dket{\psi^h_0}\}_h$ and backward propagate the result to time $t_n$,
\begin{equation}
\dket{\phi_{n}^h}=U_{n+1}^\dagger \cdots U_{N}^\dagger U_T\dket{\psi_0^h}.
\end{equation}

Equations \eqref{stgradient} and \eqref{gategradient} provide the analytical expressions needed for gradient descent. In the next two subsections, we will discuss how to numerically evaluate these expressions in a memory-efficient way.

\begin{figure*}
  \includegraphics[width=0.8\linewidth]{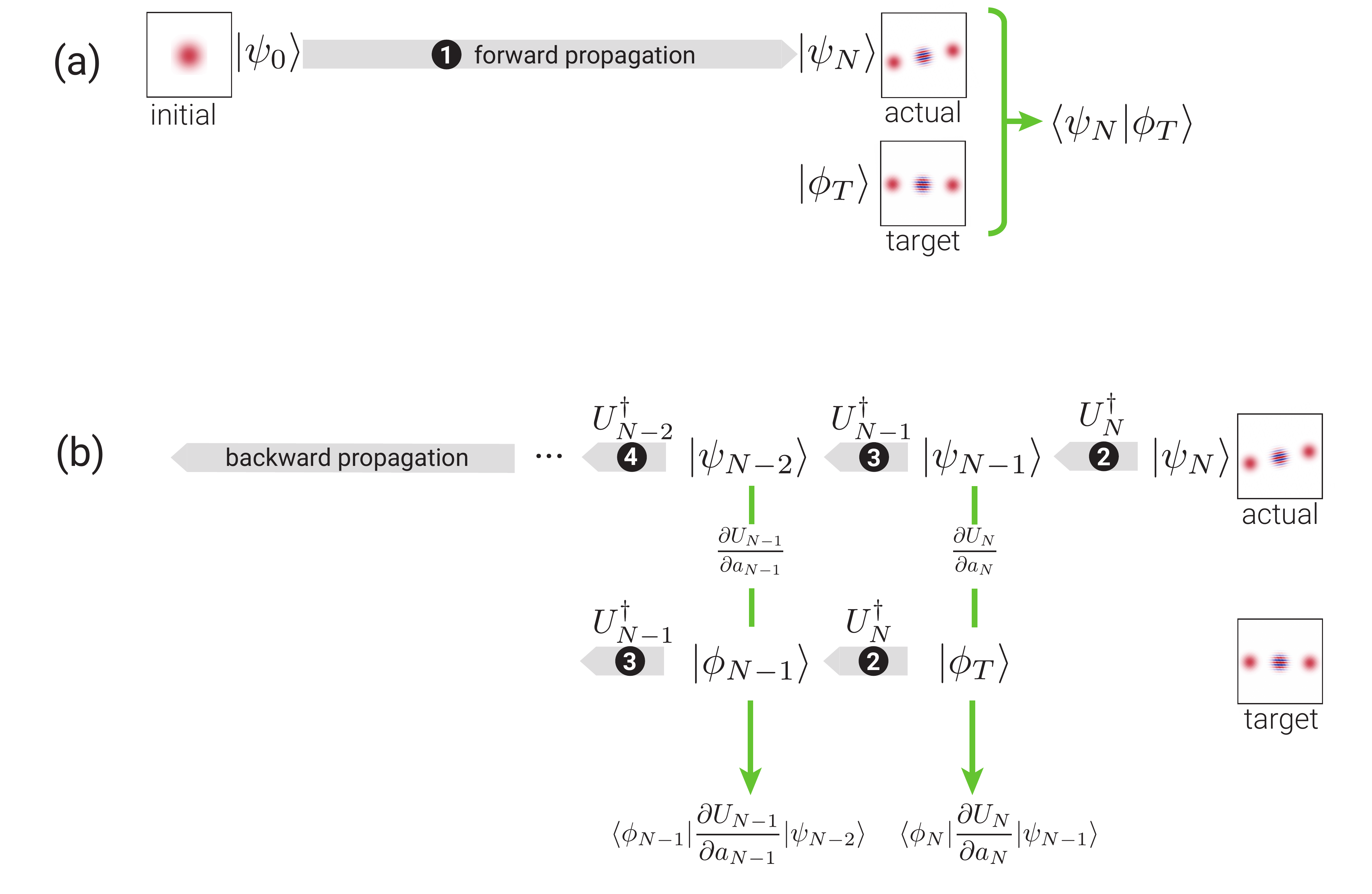}
  \caption{The forward-backward propagation scheme to compute the components of the gradient in Eq.\ \eqref{stgradient}. (a) Forward propagate the initial state by $U_N\cdots U_1$ to obtain the realized state $\dket{\psi_N}$ and calculate the overlap $\langle\psi_N|\phi_T\rangle$, where $\dket{\phi_T}$ is the target state.
    (b) Backward propagate $\dket{\psi_N}$ and $\dket{\phi_T}$ by the propagators $U_N^\dagger,U_{N-1}^\dagger,\cdots,U_1^\dagger$ iteratively, and compute $\dbra{\phi_n}\frac {\partial U_n }{\partial a_{n}}\dket{\psi_{n-1}}$ at each step step.}
    \label{fig:GRAPE}
\end{figure*}
\subsection{Numerical implementation of hard-coded gradient}
\label{implementation}

The calculation of gradients in Eqs.\ \eqref{stgradient} and \eqref{gategradient} requires numerical evaluation of expressions of the form $U_n\dket{\Psi}$  and $\frac{\partial U_n}{\partial a_n} \dket{\Psi}$. 
Here, the short-time propagator $U_n$ is given by a matrix exponential $e^A$ and $A=-iH_n\Delta t$. 

Numerically evaluating  $e^A\dket{\Psi}$ is not without challenge. As pointed out by Moler and Van Loan \cite{expmoler2003nineteen}, several methods exist, yet none of them is truly optimal. Three methods ranked highly in reference \cite{expmoler2003nineteen} are based on solving ordinary differential equations, matrix diagonalization and scaling and squaring (combined with series expansion). ODE solving, in this context, tends to be most expensive in runtime \cite{semiadgoerz2022quantum}. We mainly on focus on scaling and squaring but comment on situations where matrix diagonalization is preferable in Sec.\ \ref{Decision tree}. 

Scaling and squaring \cite{codenotti1992error} which is based on the identity
\begin{align}
\label{scaling squaring equation}
   e^A=\big(e^{A/s}\big)^s,
\end{align}
where the right hand side matrix exponential now involves the matrix $B=A/s$ which has a norm that is reduced compared to $\norm{A}$. This generally allows for a lower truncation order $m $ of the exponential series\footnote{While Chebyshev expansion has faster convergence than Taylor expansion, their numerical implementations have the same runtime scaling. Here, we focus on Taylor expansion due to its simplicity.} , 
\begin{align}
\label{eBm equation}
    e^B\approx e^B_m:=\sum_{q=0}^m\frac{ B^q}{q!}.
\end{align}
Typically, $m$ is smaller than the one required for the original $e^A$.
Our goal is to approximate the state vector $e^A \dket{\Psi}\approx (e^B_m)^s\dket{\Psi}$ where 
\begin{align}
\label{eBmpsi}
   e^B_m\dket{\Psi}&= \parenthesis{1+B+\frac{B^2}{2!}+\cdots+\frac{B^m}{m!}}\dket{\Psi}\\\nonumber
   &=\dket{\Psi}+B\dket{\Psi}+\frac{B\parenthesis{B\dket{\Psi}}}{2}+\cdots+\frac{B\parenthesis{\cdots \parenthesis{B\dket{\Psi}}}}{m!}\text{   . }
\end{align}
According to the last expression, approximation of $e^A\dket{\Psi}$ can thus proceed by repeated application of $B=A/s$ to a vector, without ever invoking matrix-matrix multiplication. This boosts runtime performance from $O(d^3)$ to $O(d^2)$.

To proceed with the evaluation of $e^A\dket{\Psi}$, the truncation order $m$ and scaling factor $s$ are determined in such a way that the error remains below the error tolerance $\tau$, 
\begin{equation}
\label{error defnition}
    \varepsilon_{A,\dket{\Psi}}(m,s) := \frac{\norm{e^A\dket{\Psi}-\llbracket (e^B_m)^s\dket{\Psi}\rrbracket}}{\norm{e^A\dket{\Psi}}}< \tau .
\end{equation}
Here, $\llbracket \bullet{}  \rrbracket$ denotes the floating point representation of $\bullet$. Since this error $\varepsilon_{A,\dket{\Psi}}$ is difficult and costly to evaluate, we instead derive an upper bound $E_A(m,s)$ [see App.\  \ref{errorupperbound}] and resort to the inequality 
\begin{equation}
\label{accuracy inequality}
    \varepsilon_{A,\dket{\Psi}}(m,s)<E_A(m,s)\leq \tau.
\end{equation}
For a given $A$, this inequality generally has multiple solutions for $m$ and $s$. As shown in App.\  \ref{errorupperbound}, the evaluation requires $ms$ matrix-vector multiplications, which are observed to dominate the runtime. Since systematic minimization of $ms$ requires significant runtime itself, we instead: (1) select solutions to Eq.\ \eqref{accuracy inequality} with the smallest $s$ which is denoted by $\mathbb{s}$, and (2) among these pairs, choose the one with smallest $m$ (which we denote by $\mathbb{m}$). Thus, calculating $e^A\dket{\Psi}$ requires
\begin{equation}
\label{mudefnition}
    \mu:=\mathbb{m}\mathbb{s}
\end{equation}
matrix-vector multiplications. As shown in App.\  \ref{numerical experienment} for two concrete examples, our method for evaluating $e^A\dket{\Psi}$ can achieve better runtime performance and accuracy than \texttt{Scipy}'s \texttt{expm\textunderscore multiply} function \cite{al2011computing} .
\begin{table*}[ht!]

\renewcommand{\arraystretch}{1.5}
\centering 
\caption{Memory usage and runtime scaling with respect to Hilbert space dimension $d$, number of time steps $N$,  stored Hamiltonian matrix elements $\kappa=\kappa(d)$ and the number of matrix-vector multiplications required for evaluating propagator-state products $\mu=\mu(d)$. (The scaling of $\mu$ and $\kappa$ depends on the specific optimization task.) The table contrasts scaling for state transfer and gate operation for hard-coded gradients (HG), full automatic differentiation (full-AD) and semi-automatic differentiation (semi-AD). } 
\begin{tabular}
{p{0.27\textwidth}p{0.17\textwidth}p{0.17\textwidth}p{0.17\textwidth}p{0.17\textwidth}} 
\hline
\hline
& \multicolumn{2}{c}{\textbf{\textsc{State transfer}}} & \multicolumn{2}{c}{\textbf{\textsc{Gate operation}}}\\
\hline 
& \textbf{Runtime} & \textbf{Memory usage} & \textbf{Runtime} & \textbf{Memory usage}
\\
\hline
Hard-coded gradient & $\Theta(N\mu \kappa)$ & $\Theta(d+\kappa)$ & $\Theta(N\mu \kappa d)$ & $\Theta(d+\kappa)$\\ 
Full automatic differentiation & $\Theta(N\mu \kappa)$ & $\Theta(N\mu d+\kappa)$ & $\Theta(N\mu \kappa d)$ & $\Theta(N\mu d^2)$  \\
Semi-automatic differentiation & $\Theta(N\mu \kappa)$ & $\Theta(Nd+\kappa)$ & $\Theta(N\mu \kappa d)$ & $\Theta(Nd^2)$\\ 
\hline \hline
\label{scalingtable}
\end{tabular} 
\label{tablescaling}
\end{table*}

Having addressed the challenges associated with numerically evaluating $e^A\dket{\Psi}$, we now turn to the calculation of $\frac{\partial U_n}{\partial a_n} \dket{\Psi}$, which is crucial for computing cost function gradients as in Eqs.\ \eqref{stgradient} and \eqref{gategradient}. We base this evaluation on the approximation 
\begin{align}
    \frac{\partial U_n}{\partial a_n} \dket{\Psi}\approx \frac{\partial (e^{B_n}_m)^s}{\partial a_n} \dket{\Psi},
\end{align}
where $B_n={-iH_n\Delta t}/{s}$. We proceed by using auxiliary matrix method based on the identity \cite{auxigoodwin2015auxiliary,najfeld1995derivatives}
\begin{equation}
\label{auxiliary equation}
    \begin{aligned}
    (e^{\mathcal{A}_n/s}_m)^s
    \binom{0}{|\Psi\rangle}&=
    \binom{
    \frac{\partial (e^{B_{n}}_m)^s}{\partial a_n}  \dket{\Psi}}{{(e^{B_n}_m)^s|\Psi\rangle}}.
\end{aligned}
\end{equation}
This relates $(\partial (e^{B_n}_m)^s / \partial a_n) \dket{\Psi}$ to the exponential of the auxiliary matrix $\mathcal{A}_n$ acting on a vector. Here, $\mathcal{A}_n$ is defined as
\begin{equation}
\mathcal{A}_n:=
{\biggl(\begin{matrix}
-iH_n \Delta t& -ih_c\Delta t \\
0 & -iH_n\Delta t
\end{matrix}\biggr)},
\end{equation}
which is a $2\times2$ matrix of operators each acting on our Hilbert space with dimension $d$. Once a basis is chosen, $\mathcal{A}_n$ can 
be represented as a $2d\times2d$ matrix of complex numbers.


\subsection{Efficient evaluation of hard-coded gradients}
\label{Computationally and memory efficient algorithm}

The analytical gradient expressions, such as $\nabla C_1^{st}$ in Eq.\ \eqref{stgradient}, are generally complicated and involve a large number of contributing quantities. Depending on grouping and ordering of operations, these quantities may need to be stored simultaneously or computed repeatedly with critical implications for runtime and memory usage. The scheme originally proposed by Khaneja\ \textit{et al.}\ \cite{nmrkhaneja2005optimal} strikes a favorable compromise that avoids the proliferation of stored intermediate state vectors. More specifically, the final quantum state $\dket{\psi_N}$ is obtained by forward propagating $\dket{\psi_0}$. Then, backward propagation of $\dket{\psi_N},\dket{\phi_T}$ is carried out to calculate the matrix element $\langle\phi_{n}|\frac {\partial U_n }{\partial a_{n}}|\psi_{n-1}\rangle$ needed for the $n$-th component of the gradient, thus building up the gradient in descending order. For both propagation directions, only the current intermediate states are stored, leading to a memory cost of $\Theta(1)$.\footnote{Here, $f(x)=\Theta
(g(x))$ is the usual Bachmann–Landau notation for $f(x)$ bounded above and below by $g(x)$ in the asymptotic sense.}


\section{Scaling analysis and benchmarking}
\label{Scaling analysis and benchmarking}
We consider the following three methods to compute gradients within GRAPE: evaluation of hard-coded gradients, full automatic differentiation\ \cite{pacleung2017speedup}, semi-automatic differentiation \cite{semiadgoerz2022quantum}. To determine the method most appropriate for a given problem, we inspect the memory usage and runtime scaling of these three approaches. Following this analysis, we present benchmark studies of specific state-transfer and gate-operation problems to illustrate the scaling behavior.

\subsection{Scaling analysis of runtime and memory usage}
\label{scaling analysis}
Runtime and memory usage can pose significant challenges when performing optimization tasks for quantum systems with large Hilbert space dimension ($d$) or for time evolution requiring significant number of time steps ($N$). In this subsection, we analyze the scaling of memory usage and runtime with respect to both $N$ and $d$, focusing on the cost contribution $C_1$ for state-transfer and unitary-gate optimization. Our findings can be generalized to other cost contributions beyond $C_1$, as elaborated in App.\  \ref{gradient}.


\paragraph{Evaluation of hard-coded gradients (HG) using forward-backward propagation.}
\label{AGscaling}

We start by considering state-transfer problems. The \textit{memory usage} for evaluating $\nabla C_1^{st}$ is mainly determined by the required storage of quantum states and the Hamiltonian. In forward-backward propagation, the number of states and instances of Hamiltonian matrices stored at any given time is of the order of unity. Each individual state vector of dimension $d$ consumes memory $\sim\Theta(d)$. Memory usage for the Hamiltonian depends on the employed storage scheme. In the ``worst" case of dense matrix storage, the memory required for Hamiltonian is $\Theta(d^2)$. For sparse matrix storage, memory usage generally depends on sparsity and is determined by the number of stored Hamiltonian matrix elements (here denoted as $\kappa$). For example, for a diagonal or tridiagonal matrix, memory usage is $\kappa\sim\Theta(d)$; for a linear chain of qubits with nearest-neighbor $\sigma_z$ coupling, it is $\Theta(d\, \log_2 d)$. Combining the scaling relations above gives us an overall asymptotic behavior of memory usage $\Theta(d+\kappa)$.

We observe that \textit{runtime} of the scheme is dominated by the required $\Theta(N)$ evaluations of propagator-state products and their derivative counterparts. Numerically approximating a single propagator-state product requires $\mu$ matrix-vector multiplications [see Eq.\ \eqref{mudefnition}], where runtime of one instance of matrix-vector multiplication scales as $\Theta(\kappa)$.  The implicit dependence of $\mu=\mu(d)$ on $d$ is specific to each particular system and can be difficult to obtain analytically. Numerically, we find for a driven transmon coupled to a cavity that $\mu\sim\Theta(\sqrt{d})$; for a linear chain of coupled qubits, the scaling is $\Theta(\log_2 d)$, see App.\  \ref{runtimescalingofUpsi} for more details. 
Once combined, the above individual scaling relations lead to an overall runtime asymptotic behavior 
$\Theta(N\mu\, \kappa)$. For example, in the case of the linear chain of coupled qubits, runtime scales as $\Theta(Nd\, (\log_2 d )^2)$.

We next turn to gate-optimization problems, where scaling behavior of memory usage and runtime are as follows. We calculate $\nabla C_1^{ g}$ by sequentially 
applying forward-backward propagation to each of the $d$ basis states. The memory usage is $\Theta(d+\kappa)$, the same as for $\nabla C_1^{st}$. The runtime scaling is $\Theta(N\mu\kappa d)$ which is $d$ times larger than the runtime scaling for $\nabla C_1^{st}$. \footnote{Another way of calculating $\nabla C_1^{ g}$ consists of parallelized forward-backward propagation of all basis states. The concrete scaling behavior then depends on the parallelization scheme; that discussion is beyond the scope of this paper.}

\paragraph{Full automatic differentiation (full-AD)}
Reverse-mode automatic differentiation extracts gradients by a forward pass and a reverse pass through the computational tree. With the forward pass the cost function $C$ is computed and the backward pass accumulates the gradient via chain rule.

The \textit{memory usage} for evaluating $\nabla C_1^{st}$ is determined by the requirement that all intermediate numerical results must be stored as part of the forward pass and act as input to the reverse pass. Specifically, the forward pass evolves the initial state vector to the final state vector in $N$ time steps. The total number of vectors that have to be stored along the way is $\Theta
(N\mu)$, since computation of a single short-time propagator-state product via scaling and squaring necessitates $\mu$ iterations. In addition, considering the memory usage for the Hamiltonian, the full scaling is $\Theta(N\mu d +\kappa)$. 

The \textit{runtime} for evaluating $\nabla C_1^{st}$ combines forward and reverse pass. It is dominated by the evaluation of $\Theta(N)$ propagator-state products, each of which has the runtime scaling $\Theta(\mu\kappa)$, leading to an overall runtime scaling of $\Theta(N\mu \kappa )$.

For gate-optimization problems, $C_1^{ g}$ is evaluated by evolving not only a single initial state but a whole set of $d$ basis vectors. Accordingly, memory usage increases by a factor of $d$ relative to state transfer, resulting in the overall memory scaling $\Theta(N\mu d^2)$. Assuming that the evolution of individual basis vectors is not parallelized, runtime also grows by a factor of $d$, so that runtime scaling for $\nabla C_1^{ g}$ is $\Theta(N\mu\kappa d)$.

\paragraph{Semi-automatic differentiation (semi-AD)}
For gradients of cost contributions that are tedious to derive analytically, semi-automatic differentiation \cite{semiadgoerz2022quantum} offers an alternative approach in which some derivatives are hard-coded and others treated by automatic differentiation. Compared to full-AD, semi-AD has the benefit that the memory usage scaling is independent of $\mu$ while maintaining the same runtime scaling. 

\paragraph{Comparison}
The results of this scaling analysis are summarized in Tab.\ \ \ref{tablescaling}. For full-AD, memory usage grows linearly with respect to $\mu$ and number of times steps $N$. By contrast, memory usage for HG does not increase with $\mu$ and $N$, but remains constant. This advantage is not bought at the cost of worse runtime scaling. Therefore, HG is preferable over full-AD in the optimal control of large quantum systems where memory usage would otherwise be prohibitively large. Semi-AD presents an interesting compromise viable whenever the memory bottleneck is not too severe.

 We illustrate the discussed scaling by benchmark studies for concrete optimization problems in the following section.

\subsection{Benchmark studies}
Our first benchmark example studies the following state-transfer problem. Consider a setup in which a driven flux-tunable transmon is capacitively coupled to a superconducting cavity, and perform a state transfer from the cavity vacuum state to the 20-photon Fock state. The Hamiltonian in the frame co-rotating with the transmon drive is
\begin{equation}
\begin{aligned}
\label{fockH}
    H_1(t)=\Delta b^\dagger b+\frac{1}{2}\alpha b^\dagger b(b^\dagger b-1)+g(cb^\dagger+c^\dagger b)\\+a_{z}(t)b^\dagger b
    +a_{x}(t)(b+b^\dagger).
\end{aligned}
\end{equation}
Here, the drive is both longitudinal and transverse,  $c$ and $b$ are the annihilation operators for cavity photons and transmon excitations respectively. $\alpha$, $g$ and $\Delta$ denote anharmonicity, interaction strength and difference between the transmon and cavity frequency. 
 
Optimization proceeds by minimizing the infidelity $C_1^{st}$ and limiting the occupation of higher transmon levels by including the cost contribution $C_2^{st}$ [Eq.\ \eqref{c2st}]. The latter has the additional benefit of enabling the truncation to only a few transmon levels. We measure runtime and peak memory usage of full-AD\footnote{AD is implemented by the package \texttt{Autograd} \cite{maclaurin2015autograd} in this benchmark. This package does not support sparse linear algebra, so we implement dense matrix storage for the Hamiltonian in both full-AD and HG for fair comparison.} and HG for a single optimization iteration. The runtime is measured by the package \texttt{Temci} \cite{temci} and
the peak memory usage is monitored by the \texttt{memory\textunderscore profiler} package \cite{memory}. The optimization is performed on an AMD Ryzen Threadripper 3970X 32-Core CPU with 3.7\,GHz frequency.

\begin{figure}
  \includegraphics[width=\linewidth]{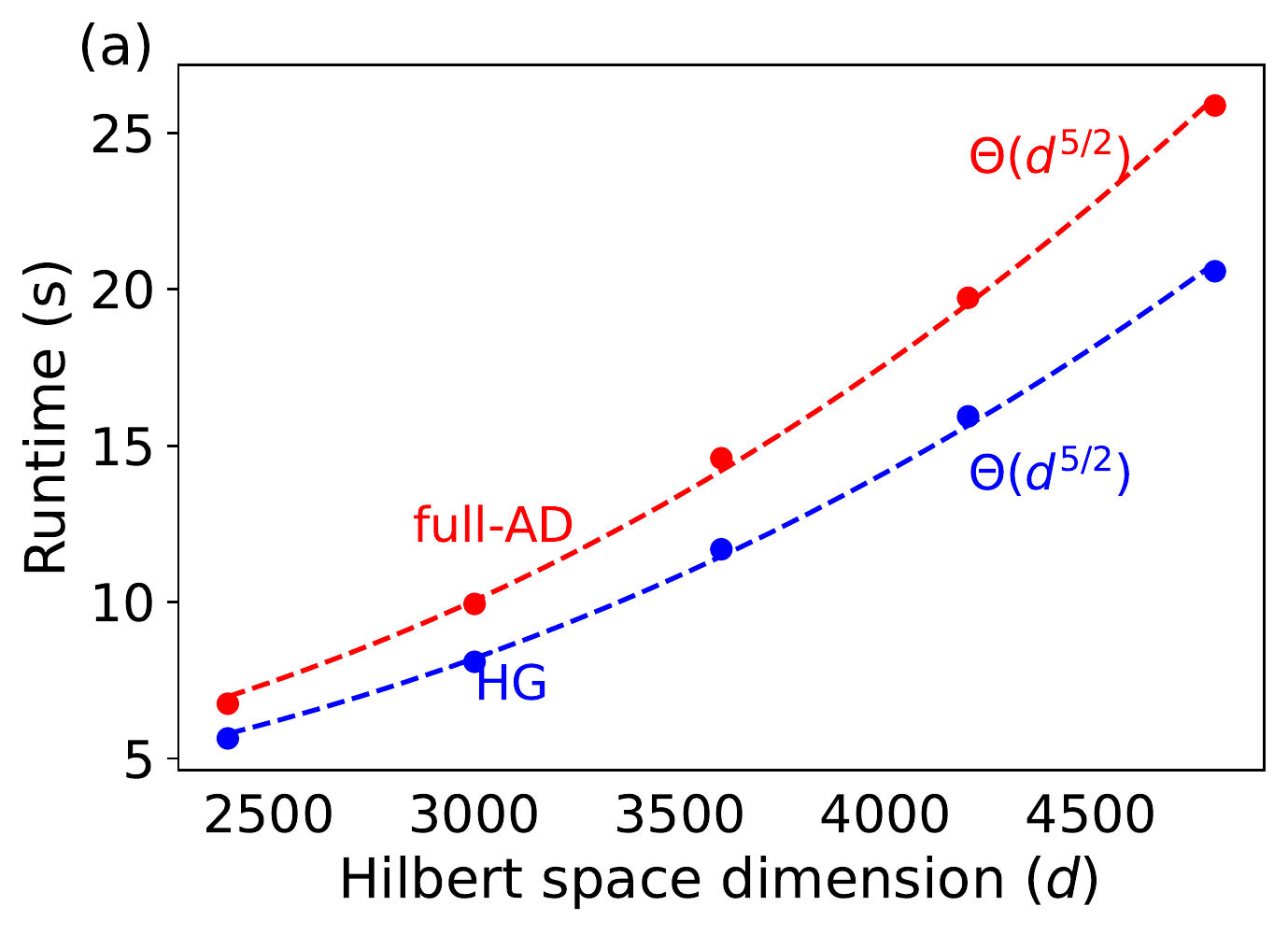} 
  \includegraphics[width=\linewidth]{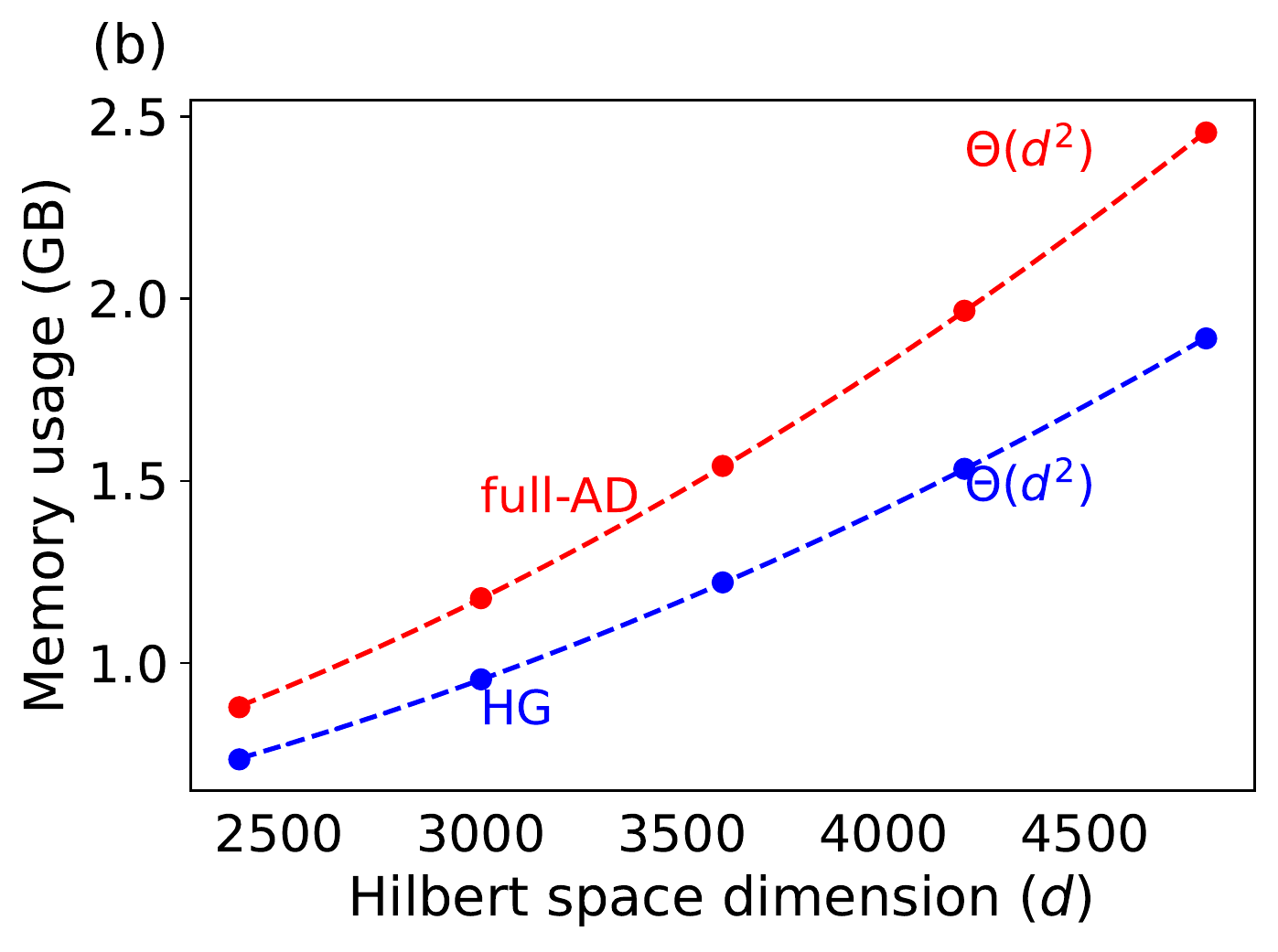}
  \includegraphics[width=\linewidth]{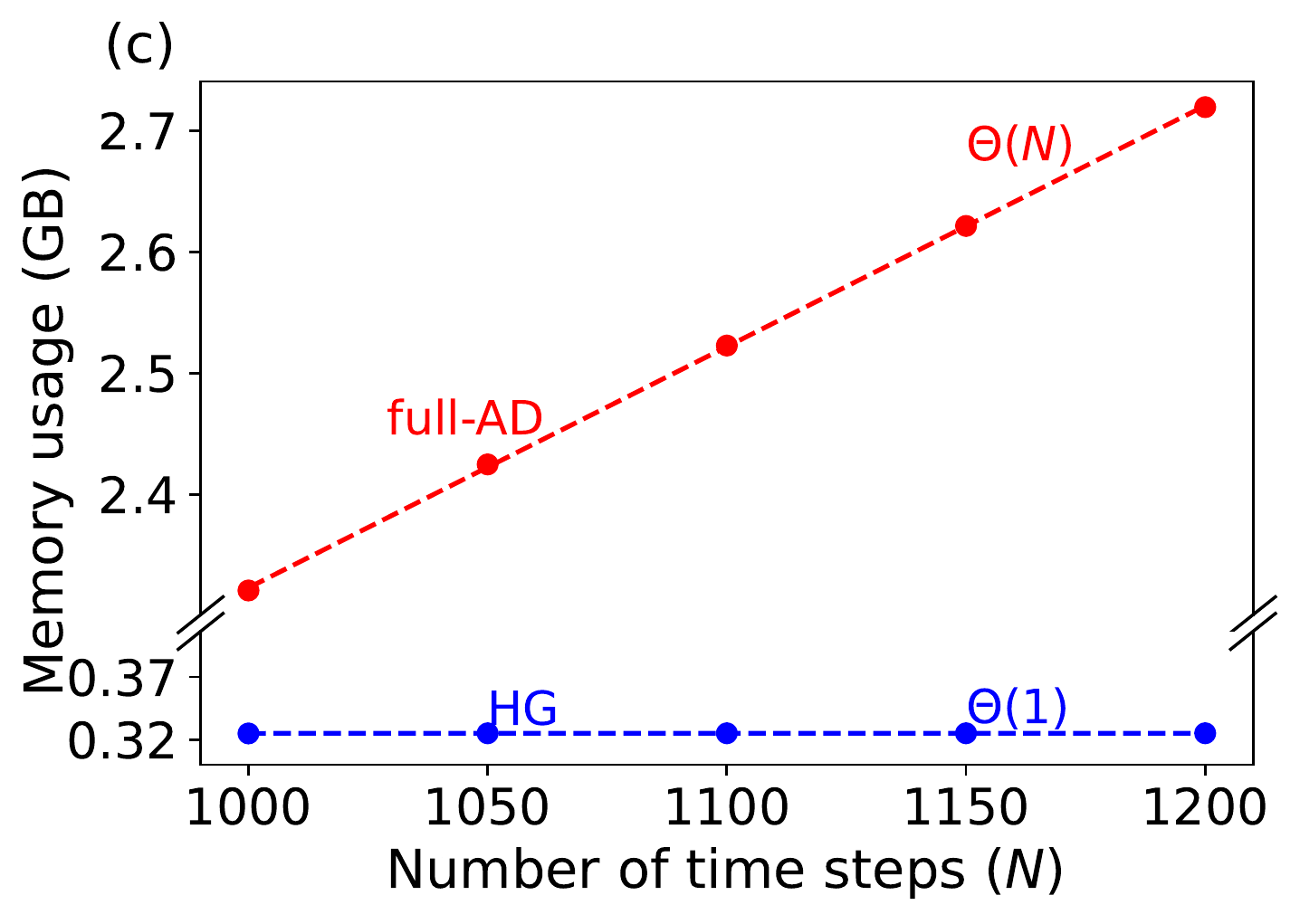}
  \caption{Benchmark results for Fock state generation in a system consisting of a cavity coupled to a transmon. (a) Runtime and (b) memory usage per evolution time step versus Hilbert space dimension $d$. The dimension is varied by increasing cavity dimension while fixing the transmon dimension to 6. (c) Memory usage as a function of the number of time steps for $d=600$. In all panels, data points are results from benchmarking and dashed lines are power law fits. (Results are obtained for: $\Delta/2\pi=3\,$GHz,  $g/2\pi=0.1\, $GHz, $\alpha/2\pi=-0.225\ $GHz and constant controls $a_z/2\pi=a_x/2\pi=0.1\ $GHz.) }
  \label{fig:state_time}
\end{figure}

\begin{figure}
  \includegraphics[width=\linewidth]{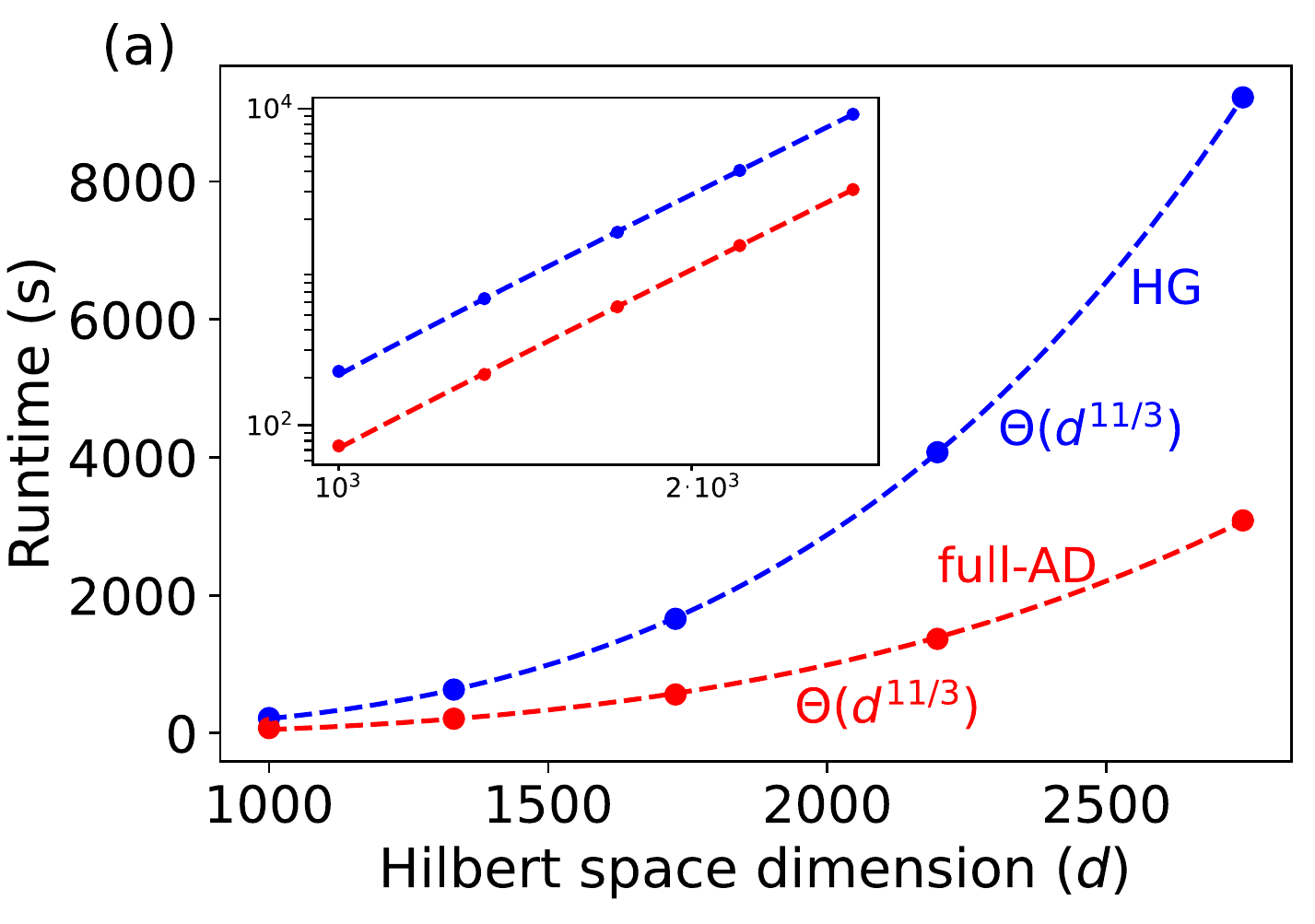}
  \includegraphics[width=\linewidth]{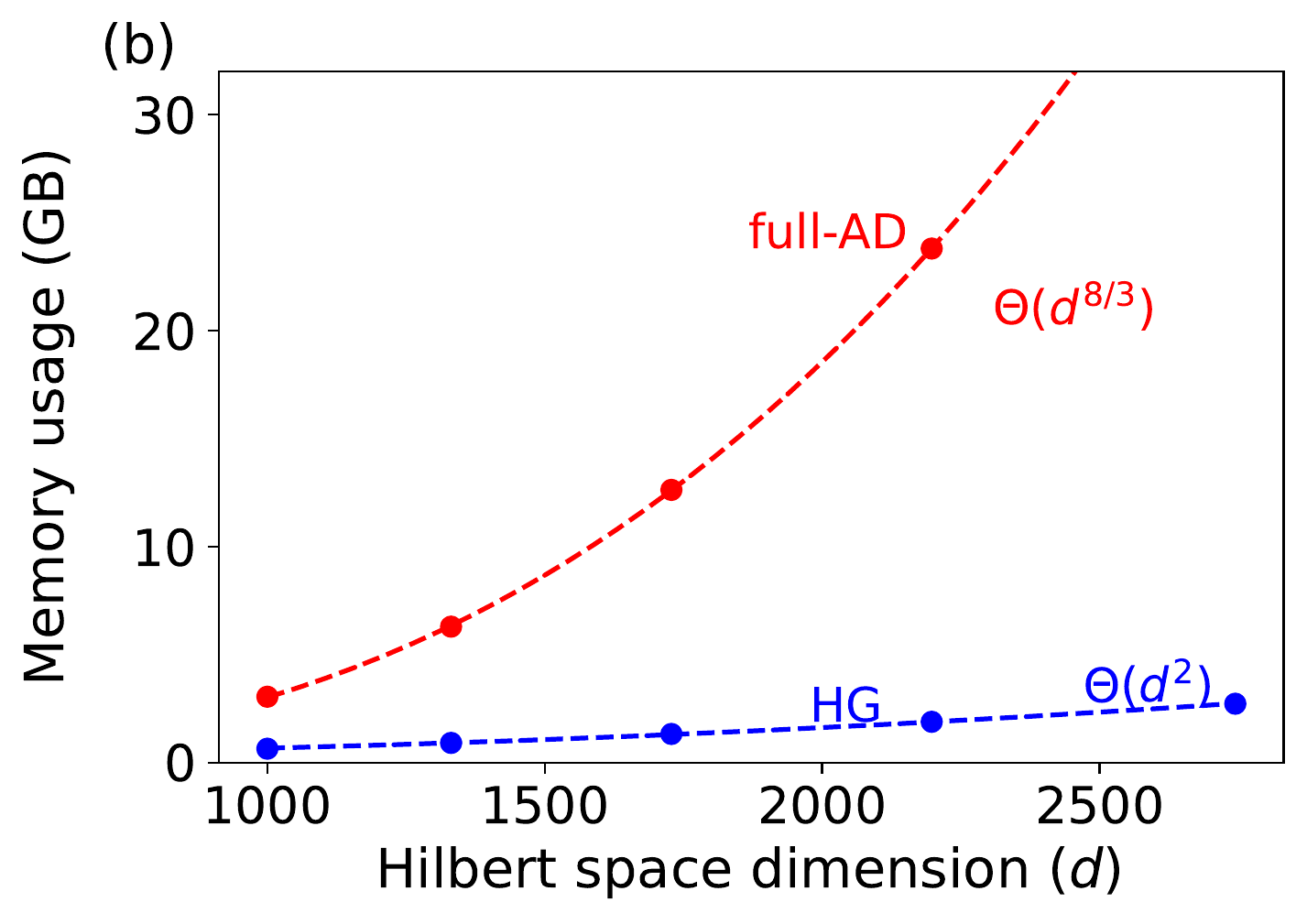}
  \caption{Benchmark results for unitary gate optimization in a system consisting of three coupled transmons. (a) Runtime and (b) memory usage per evolution time step versus Hilbert space dimension $d$. The dimension is increased by enlarging the dimension of each transmon simultaneously. The inset confirms that runtime scaling for HG and full-AD has the same power law. In all panels, data points are results from benchmarking and dashed lines are power law fits. (Results are obtained with the following parameters: $g/2\pi=$ 0.1GHz, $\alpha/2\pi=-0.225\, $GHz and $a^{(\nu)}/2\pi=0.1\ $GHz.)}
  \label{fig:gate}
\end{figure}
Benchmark results for the state-transfer optimization are shown in Fig.\ \ref{fig:state_time}. We consider a single time step ($N=1$) and measure the runtime versus $d$. The results in Fig.\ \ref{fig:state_time}(a) confirm that full-AD and HG have the same scaling of runtime per time step with respect to $d$ (Tab.\ \ref{scalingtable}). The observed runtime scaling $\Theta(d^{\frac{5}{2}})$ is consistent with $\kappa \sim \Theta(d^2)$ due to dense matrix storage for $H_1$, and
$\mu\sim \Theta(d^{\frac{1}{2}})$ (see App.\  \ref{runtimescalingofUpsi}). The scaling of required memory resources is illustrated in Fig.\ \ref{fig:state_time}(b): memory usage scales with dimension $d$ as $\Theta(d^2)$ for both full-AD and HG, consistent with $\kappa\sim \Theta(d^2)$ being the dominant contribution. As anticipated, the memory usage scaling with respect to $N$ differs between full-AD and HG [Fig.\ \ref{fig:state_time}(c)]: for full-AD, the scaling is linear with $N$; for HG, it is independent of $N$. This confirms the favorable memory efficiency of HG compared to full-AD in optimization problems with large number of time steps.

\begin{figure*}

\includegraphics[height=0.8\linewidth]{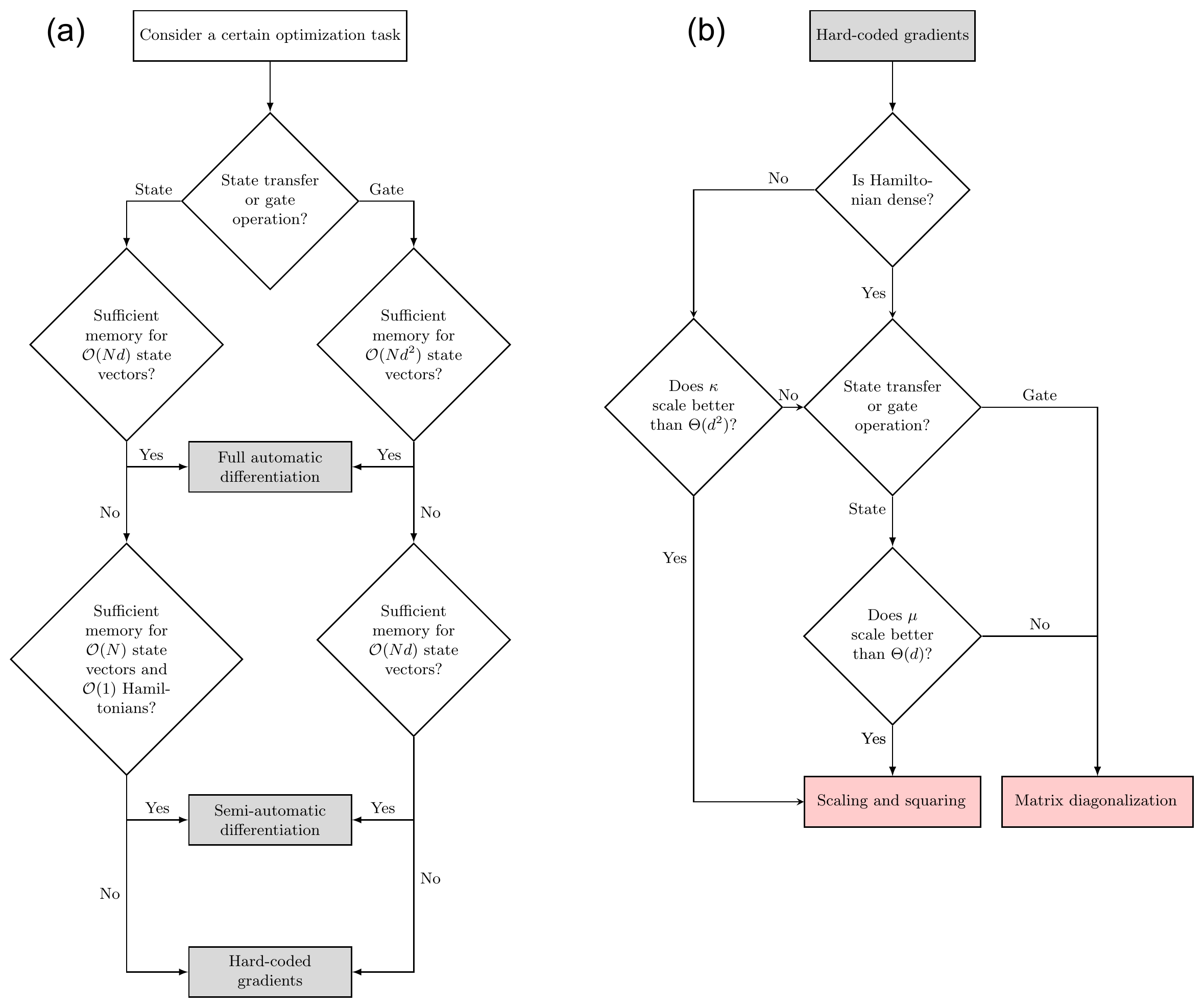}
\caption{Decision trees for selecting the optimal numerical strategy. (a) Choosing among three numerical methods for evaluating gradients: full automatic differentiation, semi-automatic differentiation, hard-coded analytical gradients. (Here, we we assume $\mu\sim\Theta(d)$; for different scaling see Tab.\ \ref{scalingtable}.) (b) Determining the appropriate strategy for evaluating propagator-state products.}

\label{fig:decision tree}
\end{figure*}

We next turn to an example for optimizing a unitary gate. This benchmark is performed for three driven transmons with nearest neighbor coupling. The target gate consists of simultaneous Hadamard operations on each of the three transmons. For concreteness, we consider the case of transmons with the same frequency, driven resonantly. The full Hamiltonian in the rotating frame of the drive is
\begin{equation}
\begin{aligned}
\label{gateH}
    H_2(t)&=\sum_{\nu=1}^{3}\bigg[\frac{1}{2}\alpha b_\nu^\dagger b_\nu (b_\nu^\dagger b_\nu-1)+a^{(\nu)}(t)(b_\nu+b_\nu^\dagger)\bigg]\\
    &\quad+g(b_1b_{2}^\dagger+b_2b_{3}^\dagger+\text{h.c.}).
\end{aligned}
\end{equation}
The goal of the optimization is to minimize the infidelity $C_2^{ g}$. The setup for this benchmark is the same as in the previous example.

Fig.\ \ref{fig:gate} records the  benchmark based on a single time step ($N=1$). The observed runtime scaling with dimension $d$ is $\Theta(d^{\frac{11}{3}})$ for both full-AD and HG, see Fig.\ \ref{fig:gate}(a). This power law arises from $\Theta(\mu\kappa d)$ for $\kappa\sim \Theta(d^2)$ and $\mu\sim \Theta(d^{\frac{2}{3}})$ [see App.\  \ref{runtimescalingofUpsi}]. The memory usage scaling with $d$ is illustrated in Fig.\ \ref{fig:gate}(b), confirming that full-AD has an unfavorable extra $\mu \sim \Theta(d^{\frac{2}{3}})$ dependence compared to HG (Tab.\  \ref{scalingtable}). Throughout our benchmarks, this causes full-AD to consume about 20 times more memory than HG. This underlines that HG is preferable whenever memory constraints become a limitation in optimization problems with large Hilbert space dimension. We emphasize that this improvement is achieved without worsening the runtime scaling\footnote{We note that the runtime scaling $\Theta(N\mu\kappa d)$ of HG via scaling and squaring can in principle exceed the scaling $O(Nd^3)$ for matrix exponentiation based on matrix diagonalization (see App.\  \ref{scalingmatrixdiag}). In this case, the latter is the better option.}. 

\section{Choosing among different optimization strategies }
\label{Decision tree}
There are several considerations which determine whether full-AD, HG or semi-AD is the optimal strategy for a given optimization problem, see the decision tree Fig.\ \ref{fig:decision tree}. Whenever the Hilbert space dimension $d$ and the number of time steps $N$ are sufficiently small, then memory usage may not be a bottleneck. In this case, full-AD may be preferable, since it eliminates the need for hard-coded gradients. If the memory cost of full-AD is not affordable but analytical gradients are tedious to compute, semi-AD may be a viable compromise, though still less memory efficient than HG. In all other cases, HG is the method of choice, since it can handle optimization with larger $N$ and $d$ compared to full-AD and semi-AD. 

The choice of implementing HG via scaling and squaring or matrix diagonalization should be informed by whether the Hamiltonian is sparse and $\kappa$ scales better than $\Theta(d^2)$. In the latter case, scaling and squaring is preferable and leads to an overall memory usage $\sim \Theta(d+\kappa)$, which is better than the memory scaling $O(d^2)$ of matrix diagonalization. By contrast, for Hamiltonians with $\kappa\sim \Theta(d^2)$, these two methods are equivalent in memory usage scaling and we are free to select the one with better runtime for a specific optimization task. While the runtime for optimization using matrix diagonalization scales as $O(d^3)$ [App.\ \ref{scalingmatrixdiag}], the runtime for optimization based on scaling and squaring differs between state-transfer and gate optimizations: (1) for state transfer, the runtime scales as $\Theta(\mu d^2)$; hence, whenever $\mu$ scales better than $\Theta(d)$, scaling and squaring wins over matrix diagonalization. (2) for gate optimization, the runtime scales as $\Theta(\mu d^3)$; hence matrix diagonalization is preferable.

\section{Conclusions \label{Conclusions}}

Motivated by the memory bottleneck problem in quantum optimal control of large systems, we have revisited and compared the memory usage and runtime scaling of multiple flavors of GRAPE: hard-coded gradients (HG), full automatic differentiation (full-AD) and semi-automatic differentiation (semi-AD). These three methods are equivalent in runtime scaling but differ characteristically in scaling of memory usage. The ranking among them is as follows, going from the most to least efficient: HG, semi-AD and full-AD. Thus HG is the preferred option when facing a memory bottleneck problem. 

We have illustrated the scaling of HG and full-AD by benchmark studies for concrete state-transfer and gate-operation optimizations, respectively. As part of this, we have presented improvements in the implementation of scaling and squaring, used to evaluate propagator-state products numerically. In the example studied, we observed a speedup compared to \texttt{Scipy}'s \texttt{expm\textunderscore multiply} function. The decision tree shown in Fig.\ \ref{fig:decision tree} summarizes the choice of appropriate numerical strategy for a given optimization problem.

\section*{Acknowledgement}
We thank Thomas Propson for fruitful discussions. This material is based upon work supported by the U.S.\  Department of Energy, Office of Science, National Quantum Information Science Research Centers, Superconducting Quantum Materials and Systems Center (SQMS) under contract number DE-AC02-07CH11359. We are grateful for the following open-source packages used in our work: \texttt{matplotlib} \cite{hunter2007matplotlib}, \texttt{numpy} \cite{harris2020array}, \texttt{scipy}  \cite{virtanen2020scipy}, \texttt{sympy}  \cite{meurer2017sympy}, \texttt{scqubits} \cite{groszkowski2021scqubits}, and 
 \texttt{qoc}  \cite{pacqoc}.
\appendix
\section{Analytical gradients}
\label{gradient}
In the main text, we exclusively focused on minimizing state-transfer or gate infidelity. In this appendix, we provide the analytical gradients for other cost contributions of interest. 

The cost contribution 
\begin{equation}
\label{c2st}
C_2^{st}=\frac{1}{N}\sum_{n'=1}^N \dbra{\psi_{n'}}\Omega \dket{\psi_{n'}}
\end{equation} 
penalizes the integrated expectation value of the Hermitian operator $\Omega$. The gradient of this cost contribution can be written as
\begin{align}
\nonumber \frac{\partial C_{2}^{st} }{\partial a_{n}}
&=\frac{1}{N}\sum_{n'=1}^N \dbra{\psi_{n'}}\Omega \frac{\partial }{\partial a_{n}}\prod_{n''=1}^{n'}U_{n''}\dket{\psi_{0}}+c.c.\\
\nonumber &=\frac{1}{N}\bigg(\sum_{n'=n}^N \prod_{n''=n+1}^{n'} \dbra{\psi_{n'}}\Omega U_{n''} \frac{\partial }{\partial a_{n}}U_n\dket{\psi_{n-1}}\bigg)+c.c.\\
\label{C2stgrad} &=\frac{2}{N}\Real (\dbra{\Phi_{n}}\frac{\partial }{\partial a_{n}}U_n \dket{\psi_{n-1}}),
\end{align}
with
\begin{equation}
    \dket{\Phi_n}:=\sum_{n'= n}^N~ U_{n+1}^\dagger \cdots U_{n'}^\dagger \Omega\dket{\psi_{n'}}
\end{equation}
The vector $\dket{\Phi_n}$ obeys the recursion relation
\begin{equation}
    \dket{\Phi_n}=U_{n+1}^\dagger \dket{\Phi_{n+1}}+\Omega \dket{\psi_n}.
\end{equation}
Due to this relation, the expression in Eq.\ \eqref{C2stgrad} can be evaluated with a forward-backward scheme analogous to the one discussed in the Sec.\ \ref{Computationally and memory efficient algorithm}.

The cost contribution $C_3^{st}=1-\frac{1}{N}\sum_{n'=1}^N \left|\left\langle\phi_{T} \mid \psi_{n'}\right\rangle\right|^{2}$ sums up infidelities from all time steps (as opposed to recording the final state infidelity only, as in $C_1^{st}$). This steers the system towards its target state more rapidly, thus helping to reduce the duration of state-transfer. Similar to Eq.\ \eqref{C2stgrad}, the gradient of $C_3^{st}$ is:
\begin{equation}
\begin{aligned}
\label{state time gradient}
  \frac{\partial C_3^{st} }{\partial a_{n}}
  &=-\frac{2}{N}\Real\big(\langle \Phi_{T,n}|\frac {\partial }{\partial a_{n}}U_n|\psi_{n-1}\rangle\big) ,\\
\end{aligned}
\end{equation}
with 
\begin{equation}
    \dket{\Phi_{T,n}}:=\sum_{n'=n}^NU_{n+1}^\dagger \cdots U_{n'}^\dagger\dket{\phi_T}\langle\phi_{T}|\psi_{n'}\rangle.
\end{equation}
The vector obeys the recursive relation
\begin{equation}
\dket{\Phi_{T,n}}=U^\dagger_{n+1}\dket{\Phi_{T,n+1}}+\dket{\phi_T}\langle\phi_T|\psi_n\rangle.
\end{equation}
This allows us to use a forward-backward scheme to evaluate the gradient expression in Eq.\ \eqref{state time gradient}.

For gate operations, the cost contribution analogous to $C_3^{st}$ is given by $C_3^{ g}=1- \sum_{n'=1}^N\big|\operatorname{tr}\big(U_T^\dagger  U_{n',1}\big)\big|^{2}/Nd^2$, where $U_{n',1}=U_{n'}\cdots U_1$ describes the evolution from time step 1 to $n'$. We obtain the gradient
\begin{equation}
  \frac{\partial C_3^{ g} }{\partial a_{n}}=
  -\frac{2}{Nd^2}\sum_{h=1}^d \Real  \big(\langle \Phi_{T,n}^h|\frac {\partial }{\partial a_{n}}U_n|\psi_{n-1}^h\rangle\big),
\end{equation}
with
\begin{equation}
\dket{\Phi_{T,n}^h}:=\sum_{n'=n}^N U_{n+1}^\dagger \cdots U_{n'}^\dagger U_T\dket{\psi_0^h} \operatorname{tr}( U_{n',1}U_T^{\dagger}).
\end{equation}
Here, $\{\dket{\psi_0^h}\}$ forms an arbitrary orthonormal basis and 
$h=1,2,\ldots,d$ enumerates the basis states. The vector $\dket{\Phi_{T,n}^h}$ obeys the recursion relation
\begin{equation}
\dket{\Phi_{T,n}^h}=U_{n+1}^\dagger\dket{\Phi_{T,n+1}^h}+U_T\dket{\psi_0^h}\operatorname{tr}( U_{n,1}U_T^{\dagger}),
\end{equation}
again enabling the evaluation of the gradient expression with a forward-backward propagation scheme.

\section{Scaling analysis}
\label{scaling analysis for other cost contributions}
For the cost contributions discussed in App.\  \ref{gradient},
we now analyze the memory usage and runtime scaling of full-AD, semi-AD and HG.
\paragraph{Scaling analysis: hard-coded gradient.}
Thanks to the similarities in the expressions and the existence of recursion relations, forward-backward schemes akin to Fig.\ \ref{fig:GRAPE} 
can also be applied to $\nabla C_2^{st}$,$\nabla C_3^{st}$ and $\nabla C_3^{g}$. As a result, we find that the scaling analysis in Sec.\ \ref{scaling analysis} leads to the same the memory usage and runtime requirement as $\nabla C_1^{st}$ and $\nabla C_2^{g}$.

\paragraph{Scaling analysis: full-automatic differentiation.}
Despite the specific differences among the expressions of cost contributions $C_2^{st}$, $C_3^{st}$ and $C_3^{g}$, we find that the scaling of runtime and memory usage is again set by the need to store intermediate vectors and repeatedly perform matrix-vector multiplications. Consequently, the scaling remains the same as for the cost contributions discussed in Sec.\ \ref{scaling analysis}


In summary, the scaling for the state-transfer and gate-operation cost contributions matches those shown in table \ref{tablescaling}. For the scaling of semi-AD, see Ref.\ \onlinecite{semiadgoerz2022quantum}.

\section{Determine the upper bound $E_A(m,s)$}
\label{errorupperbound}
The approximation of $e^A\dket{\Psi}$ in Sec.\ref{implementation} required the determination of an upper bound $E_A(m,s)$ for the error $\varepsilon_{A,\dket{\Psi}}(m,s)$ [Eq.\ \eqref{error defnition}]. In this appendix, we show how to acquire such bound.

Considering the expression for the error [Eq.\ \eqref{error defnition}], we note that the denominator is simply 1. The error $\varepsilon_{A,\dket{\Psi}}$ is bounded as follows:
\begin{align}
\nonumber    &\varepsilon_{A,\dket{\Psi}}
    =\norm{e^A\dket{\Psi}-(e^B_m)^s\dket{\Psi}+(e^B_m)^s\dket{\Psi}-\llbracket (e^B_m)^s\dket{\Psi}\rrbracket}_2\\
\nonumber    &\quad \leq \underbrace{\norm{e^A\dket{\Psi}-(e^B_m)^s\dket{\Psi}}_2}_{\varepsilon_t}+\underbrace{\norm{(e^B_m)^s\dket{\Psi}-\llbracket (e^B_m)^s\dket{\Psi}\rrbracket}_2}_{\varepsilon_r}.\\
\label{epsilonA}
\end{align}
The  error $\varepsilon_t$ arises from the truncation of the Taylor series  and rounding error $\varepsilon_r$ is due to finite-precision arithmetics. We proceed to derive the upper bounds for these two errors separately.
\subsection{Upper bound for the truncation error $\varepsilon_t$}
We rewrite the truncation error as
\begin{equation}
\begin{aligned}
    \varepsilon_t &= \norm{e^A\dket{\Psi}-(e^B-R_m(B))^s\dket{\Psi}}_2\\
    & =  \norm{\sum_{i=1}^{s}\frac{s!}{i!(s-i)!} (-R_m(B))^i(e^B)^{s-i}\dket{\Psi}}_2,
\end{aligned} 
\end{equation}
where $R_m(B)=\sum_{q=m+1}^{\infty}B^q/q!$ denotes the matrix-valued error of the truncated exponential series.
Employing the triangle inequality and submultiplicativity \cite{higham2002accuracy}, we find the bound
\begin{equation}
\begin{aligned}
    \varepsilon_t &\leq 
    \sum_{i=1}^{s}\frac{s!}{i!(s-i)!}\big(R_m(\norm{B}_2)\big)^i\norm{e^B}_2^{s-i}\norm{\dket{\Psi}}_2\\
    &=\sum_{i=1}^{s}\frac{s!}{i!(s-i)!}\big(R_m(\norm{B}_2)\big)^i.
\end{aligned}
\end{equation}
Here the second line utilizes the fact that the state is normalized and $e^B$ is unitary. 

This upper bound must be evaluated numerically to determine $m$ and $s$. However, computing the matrix 2-norm is inconvenient since it requires the use of an eigensolver. To mitigate this, we switch to the matrix 1-norm by exploiting\footnote{Generally, $\norm{B}_2\leq\sqrt{\norm{B}_\infty\norm{B}_1}$ \cite{wilkinson1994rounding} holds true for any $B$. When $B$ is anti-Hermitian, $\norm{B}_\infty=\norm{B}_1$\ leads to inequality Eq.\ \eqref{norm inequality}.}
\begin{equation}
\label{norm inequality}
    \norm{B}_2\leq\norm{B}_1.
\end{equation}
Monotonicity of $R_m$ leads us to
\begin{equation}
    \begin{aligned}
    \label{truncation}
    \varepsilon_t &\leq 
    \sum_{i=1}^{s}\frac{s!}{i!(s-i)!}\big(R_m(\norm{B}_1)\big)^i\\
    &< sR_m(\norm{B}_1)\frac{1-\big(sR_m(\norm{B}_1)\big)^s}{1-sR_m(\norm{B}_1)}.
\end{aligned}
\end{equation}

\subsection{Upper bound for the rounding error $\varepsilon_r$}
\label{rounding error upper bound}
The rounding error $\varepsilon_r$ originates from the finite precision of the floating-point representation of real numbers as well as the finite accuracy of basic arithmetic operations. The floating-point representation  $\llbracket \xi \rrbracket$ for the real number $\xi $ can be written as $\llbracket \xi \rrbracket=\xi(1+\delta)$ where $\delta=\delta(\xi)$ is the relative deviation \cite{higham2002accuracy}. Its magnitude is always smaller than machine precision, i.e. $|\delta|<u$. Similarly, the complex number $z$ obeys 
\begin{equation}
    \label{representation error}
    \llbracket z\rrbracket=z(1+\delta)
\end{equation}
with $|\delta|\leq u$. Arithmetic operations such as addition and multiplication with two complex numbers $z_1, z_2$ incur a relative deviation $\delta=\delta(z_1,z_2)$ such that 
\begin{align}
\label{operation error}
    &\llbracket z_1 \text{ + } z_2\rrbracket =(\llbracket z_1 \rrbracket+\llbracket z_2\rrbracket)(1+\delta),\ |\delta|\leq u, \\
\nonumber    &\llbracket z_1 z_2 \rrbracket=\llbracket z_1\rrbracket \llbracket z_2\rrbracket(1+\delta),\ |\delta|\leq u',
\end{align}
where $u':=2\sqrt{2}u/(1-2u)$ \cite{higham2002accuracy}. The relative deviation shown in Eqs.\ \eqref{representation error}, \eqref{operation error} generally depend on both numbers involved as well as the operation in question. In the following, we will distinguish different $\delta$ with subscript $\nu$.

Using the Eqs.\ \eqref{representation error}, \eqref{operation error}, we derive an upper bound for the rounding error in the evaluation of the dot product which is the basic arithmetic operation in evaluating $e^B_m\dket{\Psi}$ according to Eq.\ \eqref{eBmpsi}.

\subsubsection{Upper bound for the rounding error associated with evaluating the dot product}
The rounding error incurred when evaluating the dot product is given by $|\Sigma_d-\llbracket \Sigma_d\rrbracket|$ with $\Sigma_d=\boldsymbol{w^\text{T}z}$ ( $\boldsymbol{w},\boldsymbol {z} \in\mathbb{C}^{d}$). Here, $\boldsymbol{w^\text{T}}$ is a row of $B=i\delta t H/s$. We illustrate how to bound the rounding error for $d=1,2$, and give the general result subsequently. Based on Eqs.\ \eqref{representation error} and \eqref{operation error}, we have
\begin{align}
\label{c7}
    |\llbracket \Sigma_1\rrbracket-\Sigma_1|&=|\llbracket w_1  z_1\rrbracket-w_1z_1|\\
\nonumber    &<|w_1||z_1|\big|\prod_{\nu=1}^3(1+\delta_\nu)-1\big|<\gamma_3 |w_1||z_1|.
\end{align}
In the last step, we have used $\big|\prod_{\nu=1}^n(1+\delta_\nu)-1\big|<\frac{nu'}{1-nu'}$ \cite{higham2002accuracy}, and defined $\gamma_n:=\frac{nu'}{1-nu'}$. For $d=2$, the upper bound of the error is
\begin{align}
\label{c8}
     |\llbracket \Sigma_2\rrbracket-\Sigma_2|&=|(\llbracket \Sigma_1\rrbracket+\llbracket w_2 z_2\rrbracket )(1+\delta_1)-\Sigma_2|\\
\nonumber     &<\gamma_4\sum_{i=1}^2|w_i||z_i|.
\end{align}
The general upper bound for arbitrary $d$ is given by:
\begin{equation}
    |\llbracket \Sigma_d\rrbracket-\Sigma_d| <\gamma_{d+2}\sum_{j=1}^{d}|w_j|| z_j|.
\end{equation}
Since $\boldsymbol{w^\text{T}}$ is proportional to a row of the Hamiltonian matrix, sparsity of $H$ implies sparsity of $\boldsymbol{w^\text{T}}$. Vanishing entries in $\boldsymbol{w^\text{T}}$ is special because they do not incur rounding errors. Thus, one can obtain a stronger upper bound
\begin{equation}
\label{C10}
    |\llbracket \Sigma_d\rrbracket-\Sigma_d|<\gamma_{\sigma+2}\sum_{j=1}^{d}|w_j|| z_j|
\end{equation}
where $\sigma$ is the number of non-zero elements in $\boldsymbol{w^\text{T}}$.
\subsubsection{Upper bound for the rounding error of  $e^B_m\dket{\Psi}$}
The vector $e^B_m\dket{\Psi}$ is evaluated recursively via
\begin{equation}
\label{recursive relation 1}
    b_j= \frac{B}{j}b_{j-1},\qquad e^B_j b_0=b_{j-1}+b_j
\end{equation}
where $j=1,2,\cdots,m$ and $b_0$ is a vector representation of $\dket{\Psi}$ under a choice of orthonormal basis. Before evaluating the upper bound for the error $\norm{\llbracket e^B_mb_0\rrbracket-e^B_mb_0}_2$, we first derive an upper bound for the error of an vector component $\big|\llbracket e^B_mb_0 \rrbracket^{(i)}-(e^B_mb_0) ^{(i)}\big|$. We illustrate how to bound this error for $m=0,1$, and then give the general result.

For the case $m=0$, we have the bound
\begin{equation}
    \big|\llbracket e^B_0b_0 \rrbracket^{(i)} -(e^B_0b_0)^{(i)} \big|
    =\big|\llbracket b_0\rrbracket ^{(i)} -b_0 ^{(i)}\big|<\gamma_2|b_0^{(i)}|.
\end{equation}
We then obtain the upper bound of the error for the example of $m=1$:
\begin{align}
\label{c17}
    &|\llbracket e^B_1b_0 \rrbracket^{(i)}-(e^B_1b_0 )^{(i)}|\\
\nonumber    &\quad =\big|\big(\llbracket b_0\rrbracket ^{(i)}+\llbracket 
    b_1\rrbracket^{(i)}\big)(1+\delta_1)-(e^B_1b_0 )^{(i)}\big| \\
\nonumber    &\quad=\big|\llbracket b_0\rrbracket ^{(i)}(1+\delta_1)+\frac{\llbracket Ab_0\rrbracket}{s}^{(i)}\prod_{\nu=1}^3(1+\delta_\nu)^3-(e^B_1b_0 )^{(i)}\big|\\
\nonumber    &\quad \overset{\eqref{C10}}{<}\gamma_3|b_0^{(i)}| + \gamma_{\sigma_i+5}\sum_l|B^{(il)}||b_0^{(l)}|,
\end{align}
where $\sigma_i$ is number of non-zero elements in $i$th row of the matrix $B$.
Here, the third line arises from the division error between a complex number and a real number, i.e. $\llbracket z/\xi \rrbracket=z/\xi(1+\delta)$, $|\delta|<u$  \cite{higham2002accuracy}. 
The general upper bound for arbitrary $m$ is given by:
\begin{equation}
\label{c16}
    \big|\llbracket e^B_mb_0 \rrbracket^{(i)} -(e^B_mb_0 )^{(i)}\big|<\sum_{k=0}^m\gamma_{k(\sigma_i+2)+m+2}\big(\frac{|B|^k}{k!}|b_0|\big)^i,
\end{equation}
where $|B|,|b_0|$ denote the matrix and the vector with elements $|B^{il}|$, $|b_0^{i}|$, respectively. Defining
\begin{equation}
\label{sigmapdef}
    \sigma':=\max_i \sigma_i\
\end{equation} 
and using Eq.\ \eqref{c16}, we obtain
\begin{align}
\nonumber    & \norm{\llbracket e^B_mb_0 \rrbracket -(e^B_mb_0 )}_2=\big|\big||\llbracket e^B_mb_0 \rrbracket -(e^B_mb_0 )|\big|\big|_2\\
\nonumber    
&\quad <\sum_{k=0}^m\gamma_{k(\sigma'+2)+m+2}\big|\big|\frac{|B|^k}{k!}|b_0|\big|\big|_2\\
\nonumber    
&\quad <\sum_{k=0}^m\gamma_{k(\sigma'+2)+m+2}\frac{\big|\big||B|\big|\big|_2^k}{k!}\big|\big||b_0|\big|\big|_2\\
\nonumber    &
\quad \overset{\eqref{norm inequality}}{<}\sum_{k=0}^m\gamma_{k(\sigma'+2)+m+2}\frac{\big|\big||B|\big|\big|_1^k}{k!}\big|\big||b_0|\big|\big|_2\\
 \label{B23}   &\quad =\beta||b_0||_2
\end{align}
with $\beta:=\sum_{k=0}^m\gamma_{k(\sigma'+2)+m+2}\frac{||B||_1^k}{k!}$. Here, $\beta||b_0||_2$ the desired upper bound for the rounding error of $e^B_m\dket{\Psi}$. 

\subsubsection{Upper bound for the error $\varepsilon_r$}
\label{Upper bound of }
The vector $(e^B_m)^sb_0 $ is evaluated by the recursive relation.
\begin{equation}
\label{recursive relation2}
    c_l=e^B_m c_{l-1},\qquad l=1,2,\cdots,s,
\end{equation}
where $c_0=b_0$. 
We illustrate how to bound the error $\varepsilon_r=\norm{\llbracket (e^B_m)^sb_0\rrbracket-(e^B_m)^sb_0 }_2$ for $s=1,2$, and give the general result subsequently. 

For $s=1$, we have 
\begin{equation}
\label{s1}
    \norm{\llbracket e^B_mb_0\rrbracket-e^B_mb_0 }_2\overset{\eqref{B23}}{<}\beta||b_0||_2=\beta.
\end{equation}
For $s=2$, we obtain
\begin{align}
\label{C22}
\nonumber    &\norm{\llbracket (e^B_m)^2b_0\rrbracket-(e^B_m)^2b_0 }_2\\
\nonumber   &\, \leq\norm{\llbracket (e^B_m)^2b_0\rrbracket-     
    e^B_m\llbracket e^B_mb_0\rrbracket}_2+\norm{e^B_m\llbracket e^B_mb_0\rrbracket-(e^B_m)^2b_0}_2\\
   &\, <\norm{\llbracket e^B_m \llbracket e^B_mb_0\rrbracket\rrbracket-     
    e^B_m\llbracket e^B_mb_0\rrbracket}_2+\beta\norm{e^B_m}_2,
\end{align} 
where the last step is enabled by Eq.\ \eqref{s1}.
Finally, we bound the matrix norm in the second term by
\begin{align}
\nonumber    \norm{e^B_m}_2&=\norm{e^B-R_m(B)}_2\leq 1+R_m(\norm{B}_2)\\
    &\overset{\eqref{norm inequality}}{\leq} 1+R_m(\norm{B}_1)=:\alpha.
\end{align}
Since Eq.\ \eqref{B23} holds for any complex vector $b_0$, the error in Eq.\ \eqref{C22} is further bounded by:
\begin{align}
    &\norm{\llbracket (e^B_m)^2b_0\rrbracket-(e^B_m)^2b_0 }_2\overset{\eqref{B23}}{<}\beta\norm{\llbracket e^B_mb_0\rrbracket}_2+\beta\alpha\\
\nonumber &\quad =\beta\norm{\llbracket e^B_mb_0\rrbracket-e^B_mb_0+e^B_mb_0}_2+\beta\alpha\\
\nonumber        &\quad \overset{\eqref{s1}}{<}\beta(\beta+\alpha)+\beta\alpha.
\end{align}
Generalizing this strategy further, we obtain the expression for the upper bound of $\varepsilon_r$: 
\begin{align}
\label{rouding}
\nonumber    &\varepsilon_r=\norm{\llbracket (e^B_m)^sb_0\rrbracket-(e^B_m)^sb_0 }_2\\
\nonumber   &\quad \leq \sum_{l=0}^{s-1}\norm{(e^B_m)^l \llbracket (e^B_m)^{s-l}b_0\rrbracket-(e^B_m)^{l+1}\llbracket (e^B_m)^{s-l-1}b_0\rrbracket}_2\\
    &\quad<\beta\sum_{l=0}^{s-1}(\alpha+\beta)^{s-l-1}\alpha^{l}=(\alpha+\beta)^s-\alpha^s.
\end{align}    

Combining the upper bounds for $\varepsilon_r$ and $\varepsilon_t$, we return to Eq.\ \eqref{epsilonA} and obtain
\begin{align}
\label{overallupperbound}
\varepsilon_{A,\dket{\Psi}}<\nonumber E_A(m,s)&=sR_m(\norm{B}_1)\frac{1-\big(sR_m(\norm{B}_1)\big)^s}{1-sR_m(\norm{B}_1)}\\
    &\hspace{0.4cm}+(\alpha+\beta)^s-\alpha^s,
\end{align}
where $\norm{B}_1=\norm{A}_1/s$, $\alpha=\alpha(\norm{A}_1,m,s)$ and $\beta=\beta(\sigma',\norm{A}_1,m,s)$.
\section{Numerical comparison between scaling and squaring implementations}
\label{numerical experienment}
\begin{figure*}
  \centering
  \includegraphics[width=\linewidth]{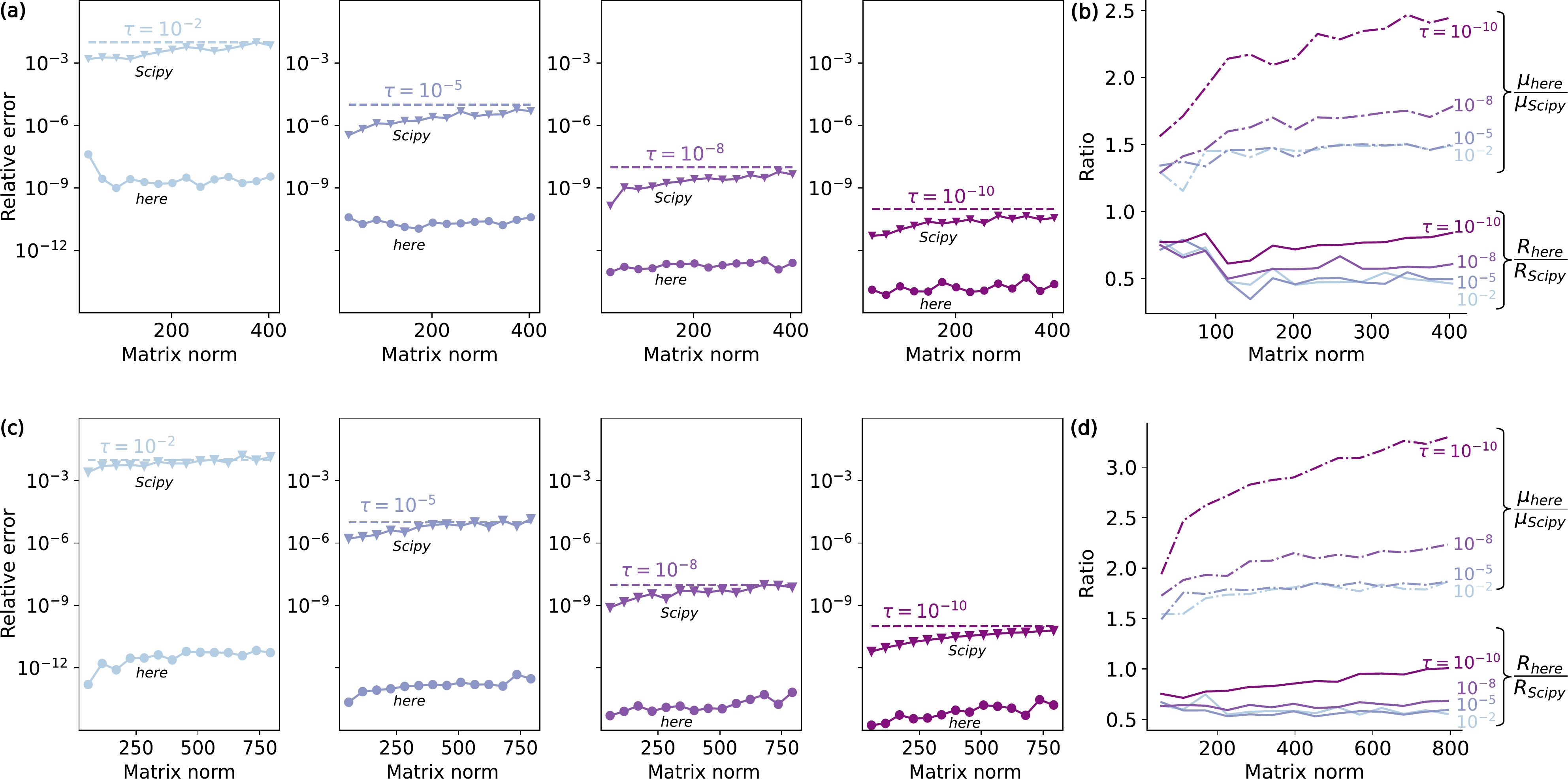}
  \caption{Comparison between scaling and squaring as implemented here and \texttt{Scipy}'s \texttt{expm\textunderscore multiply}, considering relative error and runtime. (a), (c) Relative error versus matrix norm. (b), (d) Ratios of runtime and $\mu$ (i.e. number of matrix-vector multiplications) between our  and \texttt{Scipy}'s implementations as functions of the matrix norm. The matrix norm $\norm{-iH\Delta t}_2$ is varied by increasing $\Delta t\in [1,10]$. In all panels, colors encode different error tolerances $\tau$. The results of the first and second row are obtained based on cavity-transmon and three coupled transmons Hamiltonians, respectively. The system parameters of these two Hamiltonian are the same as the ones given in the captions of Figs.\ \ref{fig:state_time}, \ref{fig:gate}, respectively.  }
  \label{fig:sas numerical experiments}
\end{figure*}
Scaling and squaring is used to evaluate the propagator-state product, $e^{-iH\Delta t }\dket{\Psi}$. We compare relative error and runtime between our implementation of scaling and squaring and \texttt{Scipy}'s implementation, \texttt{expm\textunderscore multiply}  \cite{al2011computing}. The relative errors for our implementation and \texttt{Scipy}'s implementation follow the general expression in Eq.\ \eqref{error defnition}, but differ in specific choice of $m$ and $s$. As a proxy for the exact propagator-state product, we apply Taylor expansion to matrix exponential with 400 digits precision by \texttt{Sympy} and the result of the product converges at 30 digits after decimal point. The runtime is measured with the help of the \texttt{Temci} package  \cite{temci}.

As the basis for this comparison, we use the examples of Hamiltonians \eqref{fockH} and \eqref{gateH} and implement sparse storage for them. The relative errors are shown in Fig.\ \ref{fig:sas numerical experiments}(a) and (c) as a function  versus the matrix norm $\norm{-iH\Delta t}_2$ respectively. In both examples, the relative errors for our implementation are bounded by the error tolerance $\tau$, while the relative errors for \texttt{Scipy}'s implementation tend to exceed the tolerance for large norms. There are two reasons for this tendency. First, rounding errors are not considered in the derivation of the upper bound for the errors \cite{al2011computing}; second the truncation of $e^B_m$ [Eq.\ \eqref{eBm equation}] is merely based on a heuristic criterion. 

Even though the relative errors of \texttt{Scipy}'s implementation are worse, it gains smaller number of matrix-vector multiplications, which is denoted by $\mu$. This is illustrated in Fig.\ \ref{fig:sas numerical experiments}(b) and (d), that the ratio of $\mu$ between our implementation and \texttt{Scipy}'s implementation is always larger than 1. Although \texttt{Scipy}'s implementation spends less time on matrix-vector multiplications, the total runtime of \texttt{Scipy}'s implementation is larger than the one of our implementation, see Fig. \ref{fig:sas numerical experiments}(b) and (d). This is due to the larger runtime overhead of \texttt{Scipy}'s implementation for determining $(m,s)$. 

In conclusion, our implementation has the better numerical stability and runtime than \texttt{Scipy}'s implementation in our numerical experiments. 

\begin{figure}
  \includegraphics[width=\linewidth]{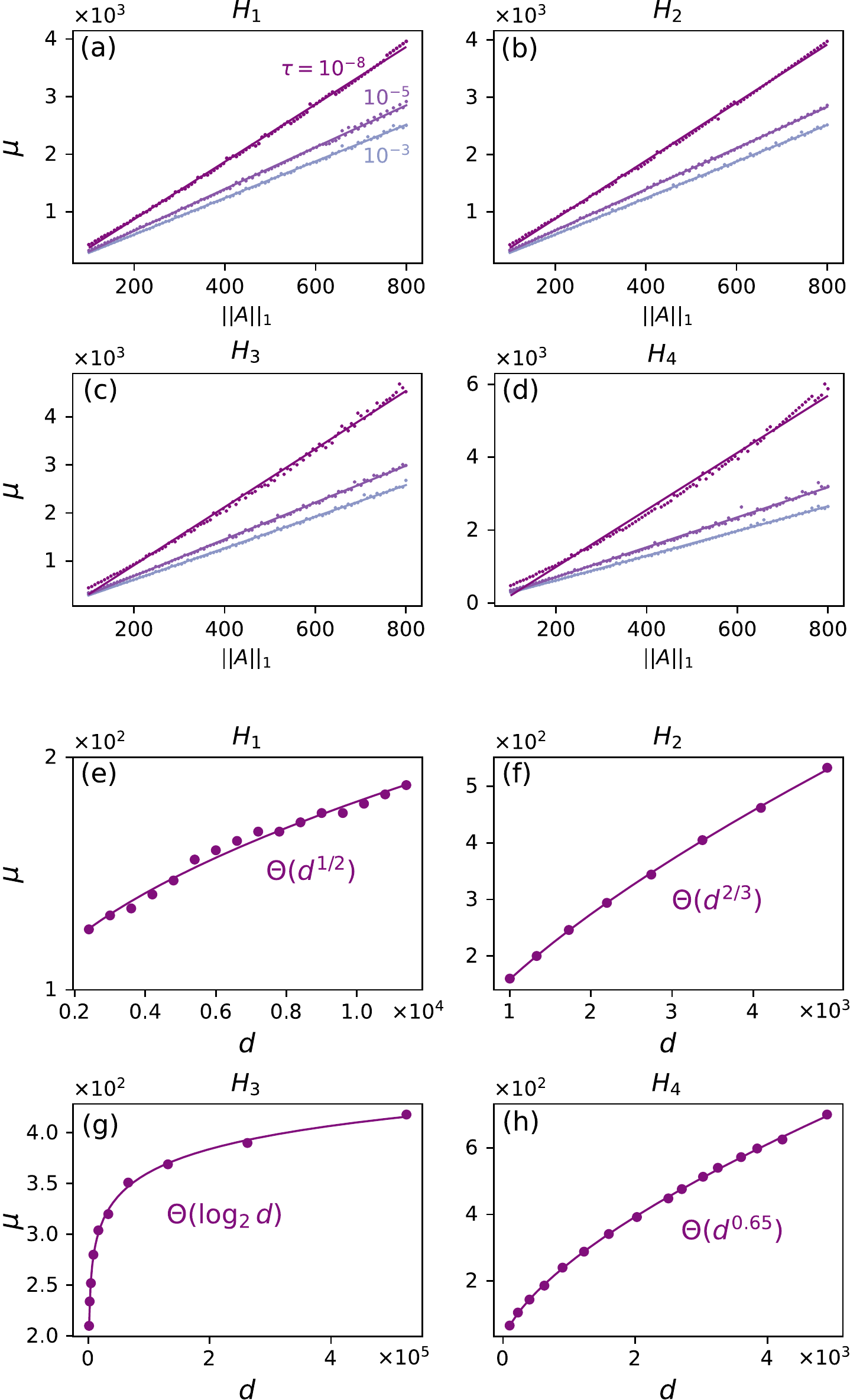}

  \caption{Scaling of $\mu$, the number of matrix-vector multiplications. (a)-(d) The scaling of $\mu$ with $\norm{A}_1$ for Hamiltonians $H_j$, $j=1,2,3,4$, when considering several error tolerances $\tau$. (e)-(h) The scaling of $\mu$ with dimension $d$ when considering $\tau=10^{-8}$ for each $H_j$, respectively.  In all figures, dots are sampling data and solid lines are fitting curves. The parameters setup of $H_1,\ H_2$ are the same as the ones given in the captions of Figs.\ \ref{fig:state_time}, \ref{fig:gate}. Parameters of $H_3$ and $H_4$ are as follows: for $H_3$, $E_{Ci}/2\pi=2.5$\ GHz, $E_{Ji}/2\pi=8.9$\ GHz, $E_{Li}/2\pi=0.5$\ GHz, $g/2\pi=0.1\ $GHz and $\phi=0.33$ ; for $H_4$, $a_x^{(\nu)}/2\pi=a_y^{(\nu)}/2\pi=0.5$\ GHz and $g/2\pi=0.1$\ GHz. }
  \label{fig:muscaling}
\end{figure}

\section{The scaling of $\mu$}
\label{runtimescalingofUpsi}
The asymptotic behavior of runtime, discussed in the main text, is governed by the scaling of $\mu$ \eqref{mudefnition} with dimension $d$ of the Hilbert space. The dependence $\mu(d)$ is specific to the matrix $A$ to be exponentiated (which is proportional to the Hamiltonian). This together with the complexity of the minimization constraint \todo{(15)} prevents us from stating a general scaling relation. We thus turn to a discussion of several concrete examples. 

 We first consider the example of the transmon-cavity system, see Eq.\ \eqref{fockH}, where we increase $d$ via the cavity dimension $d_c$ while leaving the transmon cutoff fixed. The corresponding Hamiltonian $H_1$, written in the bare product basis, then has a constant maximum number of non-zero elements in each row. (I.e., the parameter $\sigma'$ \eqref{sigmapdef} is independent of $d$.) Numerically, we find in this case that $\mu$ scales linearly with $\norm{A}_1$ \big($\mu\sim\Theta(\norm{A}_1)$\big), and this linearity holds for a wide range of $\tau$, see Fig.\ \ref{fig:muscaling}(a). In this example, two ingredients lead to the scaling $\norm{A}_1\sim \Theta(d^{\frac{1}{2}})$ : (1) $d$ is a constant multiple of $d_c$, and (2) the cavity ladder operators in $H_1$ which has one-norm scaling $\Theta(d_c^\frac{1}{2})$. Thus, we obtain the overall scaling $\mu\sim \Theta(d^{\frac{1}{2}})$, see Fig.\ \ref{fig:muscaling}(e). In the second example, we consider the system three coupled transmons, see Eq.\ \eqref{gateH}. The Hamiltonian $H_2$ is written in the harmonic oscillator basis and has a constant $\sigma'$, which leads to $\mu\sim\Theta(\norm{A}_1)$ [Fig.\ \ref{fig:muscaling}(b)]. One-norm of operator $(b^\dagger b)^2$ in $H_2$ has quadratic scaling with each transmon dimension. Since we increase $d$ by enlarging the dimensions of all three transmons simultaneously, we obtain $\norm{A}_1\sim\Theta(d^{\frac{2}{3}})$. This scaling results in $\mu\sim \Theta(d^{\frac{2}{3}})$, see Fig.\ \ref{fig:muscaling} (f).

For other Hamiltonian matrices where $\sigma'$ depends on $d$, the scaling $\mu\sim\Theta(\norm{A}_1)$ may still hold. In the following, we show two concrete examples that one where linearity holds, and another where it breaks. We first consider a driven system of $N_q$ qubits with nearest-neighbor $\sigma_z$ coupling. In the rotating frame, the system is described by 
\begin{align*}
 H_3(t)=\sum_{\nu=1}^{N_q}[a_x^{(\nu)}(t)\sigma^{(\nu)}_x+a_y^{(\nu)}(t)\sigma^{(\nu)}_y]+\sum_{\nu=1}^{N_q-1}g\sigma_z^{(\nu)}\sigma_z^{{(\nu+1)}}.
\end{align*}
Each qubit is controlled via drive fields $a_x^{(n)}$ and $a_y^{(n)}$. 
For this example, the numerical results show that scaling $\mu\sim\Theta(\norm{A}_1)$ still holds, see Fig.\ \ref{fig:muscaling}(c). Since we increase $d$ by enlarging $N_q$, we obtain $\norm{A}_1\sim\Theta(N_q)\sim\Theta(\log_2d)$. This scaling results in $\mu\sim \Theta(\log_2d)$, see Fig.\ \ref{fig:muscaling}(g). For the second example, we consider two capacitively coupled fluxonium qubits with Hamiltonian
\begin{align}
\nonumber    H_4(t)&=\sum_{i=1}^2 \big(4E_{Ci}n_i^2-E_{Ji}\cos(\varphi_i)+\frac{E_{Li}}{2}[\varphi_i-\phi_i(t)]^2\big)\\
\label{fluxonium}    &\quad + gn_1n_2.
\end{align}
Here, $E_{Ci}$, $E_{Ji}$, $E_{Li}$ denote the charging, inductive and Josephson energies of two fluxonium qubits, and $g$ is the interaction strength. $\varphi$ and $n$ are the phase and charge number operators, defined in the usual way. We represent the Hamiltonian in the harmonic oscillator basis. Fig.\ \ref{fig:muscaling}(d) shows that scaling of $\mu$ is superlinear with $\norm{A}_1$ for $\tau=10^{-8}$. We thus settle for a numerical determination of the scaling relation between $\mu$ and $d$ for $\tau=10^{-8}$ by increasing the dimensional cutoff for both fluxonium qubits simultaneously. The observed scaling is $\mu\sim \Theta (d^{0.65})$, see Fig.\ \ref{fig:muscaling}(h).

\section{Alternative numerical implementation of HG scheme and the corresponding scaling analysis}
\label{scalingmatrixdiag}
The numerical implementation of the hard-coded gradient scheme requires evaluation of the short-time propagator and its derivative. Under certain conditions, it turns out to be advantageous to numerically evaluate the short-time propagator by employing matrix diagonalization. In this appendix, we first discuss the method to compute propagator derivative and then analyze the scaling of HG.

\subsection{Evaluation of propagator derivative }
The operator $-iH\Delta t$ is anti-Hermitian, and hence can be diagonalized and exponentiated according to 
\begin{align}
\label{matrix dia}    &A=SDS^\dagger,\\
    &e^A=Se^DS^\dagger.
\end{align}
Here, $S$ is unitary and $D$ diagonal. As a result, the propagator derivative can be written as
\begin{align}
\label{D1}
    \frac{d e^{A}}{d a}=Se^D\frac{dD}{da}S^\dagger+\frac{dS}{da}e^DS^\dagger+Se^D\frac{dS^\dagger}{da}.
\end{align}
We apply the inverse unitary transformation to $\frac{d e^{A}}{d a}$ \eqref{D1} and obtain
\begin{align}
\label{D3}
    S^\dagger\frac{de^A}{da}S=e^D\frac{dD}{da}+S^\dagger\frac{dS}{da}e^D+e^D\frac{dS^\dagger}{da}S.
\end{align}
Based on $\frac{dS^\dagger}{da}S+S^\dagger\frac{dS}{da}=0$, Eq.\ \eqref{D3} can be rewritten as 
\begin{align}
\label{D4}
    S^\dagger \frac{de^A}{da}S=e^D\frac{dD}{da}+E\circ(S^\dagger\frac{dS}{da})
\end{align}
with $E_{ij}:=e^{D_{jj}}-e^{D_{ii}}$. Here, $\circ$ denotes the Hadamard product (element-wise multiplication). To obtain $dD/da$ and $dS/da$, we take the derivative of Eq.\ \eqref{matrix dia} and repeat the same steps as for Eqs.\ \eqref{D3} and \eqref{D4}, leading to
\begin{align}
\label{D5}
    S^\dagger\frac{dA}{da}S=\frac{dD}{da}+E'\circ(S^\dagger\frac{dS}{da}),
\end{align}
where $E'_{ij}:=D_{jj}-D_{ii}$. 
Since the diagonal elements of $E'$ are zero, it follows that
\begin{align}
\label{dD}
    \frac{dD}{da}=I\circ (S^\dagger\frac{dA}{da}S).
\end{align}
For off-diagonal elements of the operator in Eq.\ \eqref{D5}, we obtain
\begin{align}
\label{dF}
    F\circ (S^\dagger\frac{dA}{da}S)+G=S^\dagger\frac{dS}{da},
\end{align}
where we define 
\begin{align}
    &F_{ij}:=
  \begin{cases}
    \frac{1}{D_{jj}-D_{ii}},& \text{if $D_{jj}-D_{ii} \neq 0$}.\\
    0, & \text{otherwise}.
  \end{cases}\\
\nonumber  &G_{ij}:=
  \begin{cases}
    0, & \text{if $D_{jj}-D_{ii} \neq 0$}.\\
    (S^\dagger\frac{dS}{da})_{ij}, & \text{otherwise}.
\end{cases}
\end{align}
Inserting Eqs.\ \eqref{dD} and \eqref{dF} into Eq.\ \eqref{D4} gives
\begin{align}
\label{propagator der}
    \frac{de^A}{da}=S\big((e^D+E \circ F)\circ(S^\dagger \frac{dA}{da}S)\big)S^\dagger,
\end{align}
where we use $E\circ G=0$. 
\subsection{Scaling analysis }
The gradients of the contributions to the cost function (for the case of state transfer) are evaluated by the forward-backward propagation scheme. In this scheme, the storage required for states does not scale with the number of intermediate time steps. The matrices that must be stored are the system and control Hamiltonians as well as dense matrices dense matrices $S$, $E$ and $E^\prime$. Thus, the total memory usage scales as $O(d^2)$. The runtime of the scheme is dominated by $O(N)$ evaluations of the propagator and its derivative. Employing matrix diagonalization for the propagator evaluation requires a runtime scaling $\sim O(d^3)$ \cite{matridiagcomppan1999complexity}. Calculation of the propagator derivative involves matrix-matrix multiplication and Hadamard product with runtime scaling $O(d^3)$ and $O(d^2)$, respectively. Thus, the overall runtime of the HG scheme scales as $O(Nd^3)$. 

The gradients of the contributions to the gate-operation cost function are evaluated in an analogous manner, resulting in the runtime and memory scaling.

\bibliography{ref.bib}
\end{document}